\documentclass[12pt]{article}
\usepackage{axodraw}
\input epsf.tex

\newcommand{\beq}{\begin{equation}}
\newcommand{\eeq}{\end{equation}}
\newcommand{\beqa}{\begin{eqnarray}}
\newcommand{\eeqa}{\end{eqnarray}}
\newcommand{\beqar}{\begin{eqnarray*}}
\newcommand{\eeqar}{\end{eqnarray*}}

\newcommand{\be}{\beta}

\newcommand{\eps}{\epsilon}
\newcommand{\ga}{\gamma}
\newcommand{\Ga}{\Gamma}
\newcommand{\ka}{\kappa}
\newcommand{\inn}{\!\cdot\!}
\newcommand{\h}{\eta}

\newcommand{\la}{\lambda}

\newcommand{\p}{\phi}

\renewcommand{\t}{\theta}

\newcommand{\eg}{{\it e.g.,}\ }
\newcommand{\ie}{{\it i.e.,}\ }
\newcommand{\labell}[1]{\label{#1}} 
\newcommand{\reef}[1]{(\ref{#1})}
\newcommand\prt{\partial}
\newcommand\veps{\varepsilon}

\newcommand\cF{{\cal F}}
\newcommand\cA{{\cal A}}

\newcommand\cJ{{\cal J}}

\newcommand\cC{{\cal C}}

\newcommand\cI{{\cal I}}

\newcommand\bz{\bar{z}}

\newcommand\tK{{\widetilde K}}

\newcommand\tC{{\widetilde C}}

\newcommand\Tr{{\rm Tr}}

\begin{document}

\vspace*{1cm}

\begin{center}
{\bf \Large Towards extending the Chern-Simons couplings\\
 at order $O(\alpha'^2)$}

\vspace*{1cm}

{Mohammad R. Garousi and Mozhgan Mir}\\
\vspace*{1cm}
{ Department of Physics, Ferdowsi University of Mashhad,\\ P.O. Box 1436, Mashhad, Iran}
\\
\vspace{2cm}

\end{center}

\begin{abstract}
\baselineskip=18pt

Using the compatibility of the anomalous Chern-Simons couplings on D$_p$-branes  with the linear T-duality and with the antisymmetric B-field gauge transformations, some  couplings have been recently found for  $C^{(p-3)}$ at order $O(\alpha'^2)$. We examine these couplings with the S-matrix element of one RR and two antisymmetric B-field vertex operators. We find that  the S-matrix element  reproduces these couplings as well as some  other couplings. Each of them  is invariant under the linear T-duality and the B-field gauge transformations.

\end{abstract}

\vfill
\setcounter{page}{0}
\setcounter{footnote}{0}
\newpage


\section{Introduction and results} \label{intro}

The dynamics of the D-branes of type II superstring theories is well-approximated by the effective world-volume field theories which consist of the sum of   Dirac-Born-Infeld (DBI) and Chern-Simons (CS) actions. 
The DBI action  which describes the dynamics of the brane in the presence of the NSNS background fields at order ${\cal O}(\alpha'^0)$ is given by \cite{Leigh:1989jq,Bachas:1995kx}
\beqa
S_{DBI}&=&-T_p\int d^{p+1}x\,e^{-\phi}\sqrt{-\det\left(G_{ab}+B_{ab}\right)}\labell{DBI}
\eeqa
where $G_{ab}$ and $B_{ab}$ are  the pulled back of the bulk fields $G_{\mu\nu}$ and $B_{\mu\nu}$ onto the world-volume of D-brane\footnote{Our index conversion is that the Greek letters  $(\mu,\nu,\cdots)$ are  the indices of the space-time coordinates, the Latin letters $(a,d,c,\cdots)$ are the world-volume indices and the letters $(i,j,k,\cdots)$ are the normal bundle indices.}. The abelian gauge field can be added to the action as $B_{ab}\rightarrow B_{ab}+2\pi\alpha'f_{ab}$. The curvature corrections to this action has been found in \cite{Bachas:1999um} by requiring the consistency of the effective action with the $O(\alpha'^2)$ terms of the corresponding disk-level scattering amplitude \cite{Garousi:1996ad,Hashimoto:1996kf}.  The couplings of non-constant dilaton and B-field at the order  $O(\alpha'^2)$  have been found in \cite{Garousi:2009dj} by requiring the consistency of the curvature couplings with the standard rules of linear T-duality transformations, and by the scattering amplitude\footnote{The couplings of non-constant gauge field strength $f_{ab}$ to the DBI and CS actions have been considered in \cite{Wyllard:2000qe,Das:2001xy,Mukhi:2002gk}}. 

The CS part  which describes the coupling of D-branes to the RR potential at order ${\cal O}(\alpha'^0)$ is given by \cite{Polchinski:1995mt,Douglas:1995bn}
\beqa
S_{CS}&=&T_{p}\int_{M^{p+1}}e^{B}C\labell{CS2}
\eeqa
where $M^{p+1}$ represents the world volume of the D$_p$-brane, $C$ is the sum over all  RR potential forms,   and the multiplication rule is the wedge product.  The abelian gauge field can be added to the action as $B\rightarrow B+2\pi\alpha'f$. The curvature correction to this action has been found in \cite{Green:1996dd,Cheung:1997az,Minasian:1997mm}
 by requiring that the chiral anomaly on the world volume of intersecting D-branes (I-brane) cancels with the anomalous variation of the CS action. The curvature couplings at order $O(\alpha'^2)$ in  static gauge are
\beqa
\frac{\pi^2\alpha'^2T_p}{2!2!(p-3)!}\int d^{p+1}x\epsilon^{a_0\cdots a_{p}}{\cal C}^{(p-3)}_{a_4\cdots a_{p-4}}\bigg[R_{a_0a_1}{}^{ab}R_{a_2a_3ba}-R_{a_0a_1}{}^{ij}R_{a_2a_3ji}\bigg]\labell{Tf2}
\eeqa
They have been confirmed by the S-matrix calculation in \cite{Craps:1998fn,Morales:1998ux,Stefanski:1998yx}. The above  couplings have been extended in \cite{Becker:2010ij,Garousi:2010rn} to include the   B-field at the order  $O(\alpha'^2)$   by requiring them to be  consistent   with the  linear T-duality transformations. The new B-field couplings, however, are not invariant under the B-field gauge transformation. Adding some other B-field couplings which are themselves invariant under the linear T-duality, one can write the resulting couplings in a gauge invariant form \cite{Garousi:2010rn}. The gauge invariant couplings are 
\beqa
&&\frac{\pi^2\alpha'^2T_p}{2!2!(p-3)!}\int d^{p+1}x\epsilon^{a_0\cdots a_{p}}{\cal C}^{(p-3)}_{a_4\cdots a_{p-4}}\bigg[\frac{1}{2}H_{a_0a_1a,i}H_{a_2a_3}{}^{a,i}-\frac{1}{2}H_{a_0a_1i,a}H_{a_2a_3}{}^{i,a}
\bigg]\labell{Tf41new}
\eeqa
where   commas denote partial differentiation. Unlike the gravity couplings, the  B-field couplings \reef{Tf41new} are not invariant under the RR gauge transformation so one may expect that there should be some  other couplings as well.
 Since there are no gravity couplings for $C^{(p-3)}$ other than those given by \reef{Tf2}, we are  not allowed  to have any other T-duality invariant  couplings which include both gravity and B-fields. However, it is consistent  to have  couplings  which  involve only B-fields.   In this paper,  we will show that the S-matrix element of one RR and two B-field vertex operators produce the  couplings \reef{Tf41new} as well as some other T-duality invariant couplings.

The scattering amplitude of one RR potential $C^{(p-3)}$ and two gravitons at order  $O(\alpha'^2)$ has no massless open string or closed string pole. It has  only contact terms  given by \reef{Tf2} \cite{Craps:1998fn,Morales:1998ux,Stefanski:1998yx} which  are invariant under the RR gauge transformation. On the other hand, as we will see the scattering amplitude of one RR potential  $C^{(p-3)}$ and two B-fields at order  $O(\alpha'^2)$ has both open and closed string poles, as well as some contact terms. We will show that the sum of all these contributions at any order  of $\alpha'$  has the RR gauge symmetry. We are interested in this paper in the massless open string poles and the contact terms of the amplitude which dictate the appropriate  couplings on D-branes.  

It has been shown in \cite{Garousi:2010ki} that the CS action should  also include couplings  which involve linear  NSNS field.  These couplings have been found  by studying the S-matrix element of one RR and one NSNS vertex operators  at order $O(\alpha'^2)$ \cite{Garousi:1996ad}.  These couplings for $F^{(p)}$    are \cite{Garousi:2010ki}
\beqa
-\frac{\pi^2\alpha'^2T_p}{2!(p-1)!}\int d^{p+1}x\,\eps^{a_0\cdots a_p}\left({ F}^{(p)}_{ia_2\cdots a_p,a}H_{a_0a_1}{}^{a,i}-{ F}^{(p)}_{aa_2\cdots a_p,i}H_{a_0a_1}{}^{i,a}\right)\nonumber
\eeqa
 They  can be written in terms of the RR potential  as
 \beqa
\frac{\pi^2\alpha'^2T_p}{(p-1)!}\int d^{p+1}x\,\eps^{a_0\cdots a_p}\left[\frac{p-1}{3!}{ C}^{(p-1)}_{ia_3\cdots a_{p}}H_{a_0a_1a_2}{}^{,ia}{}_a+\frac{1}{2!}{ C}^{(p-1)}_{a_2\cdots a_p,i}(2H_{a_0a_1}{}^{a,i}{}_a-H_{a_0a_1}{}^{i,a}{}_a)\right]\labell{LTdual}
\eeqa
 They  are invariant under the linear T-duality transformations. The consistency of these couplings with the standard nonlinear T-duality \cite{TB,Meessen:1998qm,Bergshoeff:1995as,Bergshoeff:1996ui,Hassan:1999bv} requires some nonlinear couplings for $C^{(p-3)}$. However, because of the appearance of the transverse index in the RR potential, the  nonlinear T-duality transformations of the RR field in the absence of D-branes, \ie $\tC^{(n)}_{\mu\nu\cdots \alpha\beta}=C^{(n+1)}_{\mu\nu\cdots \alpha\beta y}+nC^{(n-1)}_{[\mu\nu\cdots \alpha}B^{}_{\beta]y}+\cdots$, produces some couplings which break the B-field gauge symmetry. On the other hand, since the contracted indices $i,a$ in the above equation are derivative indices, the nonlinear terms are invariant under linear T-duality at the level of two B-fields \cite{Garousi:2010rn}. So it is consistent with  T-duality to remove the terms which break the gauge symmetry.  Consider   then the following part of the nonlinear T-duality transformation of the RR potential: 
 \beqa
 \tC^{(p-1)}_{ia_3\cdots a_p}&=&C^{(p)}_{ia_3\cdots a_p y}+(p-2)C^{(p-2)}_{i[a_3\cdots a_{p-1}}B^{}_{a_p]y}+\cdots\nonumber\\
 \tC^{(p-1)}_{a_2\cdots a_p,i}&=&C^{(p)}_{a_2\cdots a_p y,i}+(p-1)\prt_iC^{(p-2)}_{[a_2\cdots a_{p-1}}B^{}_{a_p]y}+\cdots
 \eeqa
where dots represent higher nonlinear terms. Following \cite{Garousi:2010ki}, one finds that the consistency of the couplings \reef{LTdual} with the above T-duality   requires  the following couplings for $C^{(p-3)}$:
 \beqa
&&\pi^2\alpha'^2T_p\int d^{p+1}x\,\eps^{a_0\cdots a_p}\bigg[\frac{1}{3!2!(p-4)!}(B_{a_3a_4}+2\pi\alpha'f_{a_3a_4}){ C}^{(p-3)}_{ia_5\cdots a_{p}}H_{a_0a_1a_2}{}^{,ia}{}_a\labell{77}\\
&&\qquad\qquad\qquad\qquad+\frac{1}{2!2!(p-3)!}(B_{a_2a_3}+2\pi\alpha'f_{a_2a_3}){ C}^{(p-3)}_{a_4\cdots a_p,i}(2H_{a_0a_1}{}^{a,i}{}_a-H_{a_0a_1}{}^{i,a}{}_a)\bigg]\nonumber
\eeqa
where we have also used  the replacement $B\rightarrow B+2\pi\alpha'f$ to make the couplings gauge invariant.  The above couplings are invariant under the linear T-duality transformation at the level of two B-fields. They   produce some massless open string poles and contact terms which can be combined into  massless poles written in terms of field strength $H$. The massless pole corresponding to the couplings in the first line is   reproduced by the disk level S-matrix element of one RR and two B-fields vertex operators in which the RR potential carries one transverse index \cite{Garousi:2010bm}. We will show that the massless poles corresponding to the couplings in the second line are reproduced by  the S-matrix element in which the RR potential carries  only world volume indices.
 
The  S-matrix element still reproduces   some other T-duality  invariant couplings which are given by  
\beqa
&&\frac{\pi^2\alpha'^2T_{p}}{(p-3)!}\int d^{p+1}x\,\eps^{a_0a_1\cdots a_{p}}{\cal C}^{(p-3)}_{a_4\cdots a_{p}}\bigg[\frac{1}{2!2!}H^{aa_0a_1}{}_{,ab}H^{ba_2a_3}+\frac{1}{3!}H^{a_0a_1a_2}{}_{,ai}H^{iaa_3}\nonumber\\
&&\qquad\qquad\qquad+\frac{1}{2!2!}H^{a_0a_1a,i}{}_aH^{a_2a_3}{}_i+\frac{1}{3!}H^{a_0a_1a_2}{}_{,a}H^{aba_3}{}_{,b}+\frac{1}{3!}H^{a_0a_1a_2}{}_{,i}H^{iaa_3}{}_{,a}\labell{5}\\
&&\qquad\qquad\qquad-\frac{1}{3!} H^{a_0a_1a_2,a}{}_{ab}(B^{ba_3}+2\pi\alpha' f^{ba_3})-\frac{1}{2!2!} H^{a_0a_1b,a}{}_{a}(B^{a_2a_3}+2\pi\alpha' f^{a_2a_3}){}_{,b}\bigg]\nonumber
\eeqa
Note that the contracted indices $i,a,b$ are derivative indices, hence, the above couplings are all invariant under the linear T-duality transformation at the level of two B-fields. As in \reef{77}, we will see that the terms which include $(B+2\pi\alpha'f)$ produce   massless open string poles. Our limitation to calculate  the triple integrals that appear in the S-matrix element, does not allow us to calculate the coefficient of all such terms. However, there is no such limitation for calculating the contact terms.

 An outline of the paper is as follows: In section 2.1 we examine the calculation of  the S-matrix element of one RR and two NSNS vertex operators in superstring theory. We perform the calculation in full details for the RR potential $C^{(p-3)}$ which has only world volume indices and expand the amplitude at low energy. In section 2.2, using the couplings in \reef{Tf41new}, \reef{77} and \reef{5}, we calculate the massless open string poles and the contact terms for the scattering amplitude of one RR scalar and two B-fields.  We show that they   are reproduced exactly by string theory amplitude at order $O(\alpha'^2)$.

 \section{Scattering amplitude } \label{intro2}
 
A powerful method for finding the low energy field theory of the string theory is to compare the scattering amplitudes of the field theory with the corresponding  amplitudes in the string theory expanded at low energy.  The disk level scattering amplitude of one RR and two NSNS vertex operators, at low energy, produces both massless open string and closed string  poles as well as some contact terms. The  closed string poles dictate the supergravity couplings in the bulk and the couplings of one RR and one NSNS states on the brane. On the other hand, the open string poles and the contact terms dictate  the couplings of one RR and two B-fields on the brane in which we are interested in this paper. We shall show that the couplings in \reef{Tf41new}, \reef{77} and \reef{5} are produced  by the  scattering amplitude  at low energy. In the next  section, we calculate the string theory amplitude.

 \subsection{String theory amplitude}
 
The scattering amplitude of one RR and two NSNS states has been studied in \cite{Becker:2010ij,Garousi:2010bm} for a particular class of terms in the amplitude to confirm some part of the couplings resulting from the consistency of the CS action \reef{Tf2}  with the linear T-duality, and the couplings \reef{LTdual} with nonlinear T-duality.  In this paper, however, we are interested in finding all  couplings that string theory produces for the RR potential $C^{(p-3)}$ which carries only the world volume indices. 

In string theory, the tree level scattering amplitude of one RR and two NSNS states on the world-volume of a D$_p$-brane  is given  by the correlation function of their  corresponding vertex operators on the disk. Since the background charge of the world-sheet with topology of a disk is $Q_{\phi}=2$ one has to choose  the vertex operators in the appropriate pictures to produce the compensating charge $Q_{\phi}=-2$. One may choose the RR vertex operator in $(-1/2,-1/2)$ picture, and one of the NSNS vertex operators in $(-1,0)$ and the other one in $(0,0)$. However, in this picture the symmetry between the two NSNS is not manifest from the very beginning. After performing the correlators, one has to make more effort to rewrite the final result in a symmetric form. Alternatively, one can choose the RR vertex operator in $(-1/2,-3/2)$ picture \cite{Billo:1998vr} and the two NSNS vertex operators in $(0,0)$ picture. In this form the symmetry of the NSNS states is manifest from the beginning. We prefer to do the calculation  in the latter form. We will show that the final result, after using some identities, are independent of the choice of the picture. 

The scattering amplitude is given by the following correlation function:
\beqa
\cA&\sim&<V_{RR}^{(-1/2,-3/2)}(\veps_1^{(n)},p_1)V_{NSNS}^{(0,0)}(\veps_2,p_2)V_{NSNS}^{(0,0)}(\veps_3,p_3)>\labell{amp2}
\eeqa
Using the doubling trick \cite{Garousi:1996ad}, the vertex operators are given by\footnote{Our conversions set $\alpha'=2$ in the string theory calculations.}
\beqa
V_{RR}^{(-1/2,-3/2)}&\!\!\!\!\!=\!\!\!\!\!&(P_-H_{1(n)}M_p)^{AB}\int d^2z_1:e^{-\phi(z_1)/2}S_A(z_1)e^{ip_1\cdot X}:e^{-3\phi(\bz_1)/2}S_B(\bz_1)e^{ip_1\cdot D\cdot  X}:\nonumber\\
V_{NSNS}^{(0,0)}&\!\!\!\!\!=\!\!\!\!\!&(\veps_2\inn D)_{\mu_3\mu_4}\int d^2z_2:(\prt X^{\mu_3}+ip_2\inn\psi\psi^{\mu_3})e^{ip_2\cdot X}:(\prt X^{\mu_4}+ip_2\inn D\inn\psi\psi^{\mu_4})e^{ip_2\cdot D\cdot X}:\nonumber\\
V_{NSNS}^{(0,0)}&\!\!\!\!\!=\!\!\!\!\!&(\veps_3\inn D)_{\mu_5\mu_6}\int d^2z_3:(\prt X^{\mu_5}+ip_3\inn\psi\psi^{\mu_5})e^{ip_3\cdot X}:(\prt X^{\mu_6}+ip_3\inn\nonumber D\inn\psi\psi^{\mu_6})e^{ip_3\cdot D\cdot X}:
\eeqa
where  the indices $A,B,\cdots$ are the Dirac spinor indices and  $P_-=\frac{1}{2}(1-\gamma_{11})$ is the chiral projection operator which makes the calculation of the gamma matrices to be with the full $32\times 32$ Dirac matrices of the ten dimensions. 
 The matrix $D^{\mu}_{\nu}$ is diagonal with $+1$ in the world volume directions and $-1$ in the transverse directions, and
\beqa
H_{1(n)}&=&\frac{1}{n!}\veps_{1\mu_1\cdots\mu_{n}}\gamma^{\mu_1}\cdots\gamma^{\mu_{n}}\nonumber\\
M_p&=&\frac{\pm 1}{(p+1)!} \eps_{a_0 \cdots a_p} \ga^{a_0} \cdots \ga^{a_p}
\eeqa
where $\eps$ is the volume $(p+1)$-form of the $D_p$-brane. The polarization of the RR field is given by $\veps_1^{(n)}$ and the polarizations of the B-fields are given by $\veps_2,\, \veps_3$. The on-shell conditions  are 
\beqa
p_i\inn p_i=p_i^{\mu}(\veps_i){}_{\mu \cdots}=0,\, &{\rm for}& i=1,2,3
\eeqa
It is useful to write the matrix $D_{\mu\nu}$ and the flat metric $\eta_{\mu\nu}$ in terms of the two projection operators $N_{\mu\nu}$ and $V_{\mu\nu}$, \ie
\beqa
\eta_{\mu\nu}&=&V_{\mu\nu}+N_{\mu\nu}\nonumber\\
D_{\mu\nu}&=&V_{\mu\nu}-N_{\mu\nu}
\eeqa
The components of vectors projected into each of these subspaces $N$ and $V$ or $\eta$ and $D$ are independent objects. If 1 in the chiral projection $P_-$ produces couplings for $C^{(n)}$, then the $\gamma_{11}$ produces the couplings for $C^{(10-n)}$. Hence, we consider 1 in the chiral projection and extend the result to all RR potentials.

Choosing the above integral form of the vertex operators, one has to also divide the amplitude \reef{amp2} by the volume of $SL(2,R)$ group which is the conformal symmetry of the  upper half $z$-plane. We will remove this factor after  preforming the correlators. Moreover, the overall factor of the amplitude \reef{amp2} may  be fixed by comparing the final result with field theory.

Using the standard world-sheet propagators,
one can calculate the correlators in \reef{amp2}. The amplitude \reef{amp2}  can be written  as \cite{Garousi:2010bm}
\beqa
\cA&\sim&\frac{1}{2}(H_{1(n)}M_p)^{AB}(\veps_2\inn D)_{\mu_3\mu_4}(\veps_3\inn D)_{\mu_5\mu_6}\int d^2z_1d^2z_2d^2z_3\, (z_1-\bz_1)^{-3/4}\nonumber\\
&&\times(b_1+b_2+\cdots +b_{10})^{\mu_3\mu_4\mu_5\mu_6}_{AB}\delta^{p+1}(p_1^a+p_2^a+p_3^a)+(2\leftrightarrow 3)\labell{A}
\eeqa
where 
\beqa
(b_1)^{\mu_3\mu_4\mu_5\mu_6}_{AB}&\!\!\!\!=\!\!\!\!&<:S_A(z_1):S_B(\bz_1):>g_1^{\mu_3\mu_4\mu_5\mu_6}\nonumber\\
(b_2)^{\mu_3\mu_4\mu_5\mu_6}_{AB}&\!\!\!\!=\!\!\!\!&2(ip_2)_{\beta_3}<:S_A:S_B:\psi^{\beta_3}\psi^{\mu_3}:>g_2^{\mu_4\mu_5\mu_6}\nonumber\\
(b_3)^{\mu_3\mu_4\mu_5\mu_6}_{AB}&\!\!\!\!=\!\!\!\!&2(ip_2\inn D)_{\beta_4}<:S_A:S_B:\psi^{\beta_4}\psi^{\mu_4}:>g_3^{\mu_3\mu_5\mu_6}\nonumber\\
(b_4)^{\mu_3\mu_4\mu_5\mu_6}_{AB}&\!\!\!\!=\!\!\!\!&2(ip_2)_{\beta_3}(ip_2\inn D)_{\beta_4}<:S_A:S_B:\psi^{\beta_3}\psi^{\mu_3}:\psi^{\beta_4}\psi^{\mu_4}:>g_4^{\mu_5\mu_6}\labell{bs}\\
(b_5)^{\mu_3\mu_4\mu_5\mu_6}_{AB}&\!\!\!\!=\!\!\!\!&(ip_2)_{\beta_3}(ip_3)_{\beta_5}<:S_A:S_B:\psi^{\beta_3}\psi^{\mu_3}:\psi^{\beta_5}\psi^{\mu_5}:>g_5^{\mu_4\mu_6}\nonumber\\
(b_6)^{\mu_3\mu_4\mu_5\mu_6}_{AB}&\!\!\!\!=\!\!\!\!&2(ip_2)_{\beta_3}(ip_3\inn D)_{\beta_6}<:S_A:S_B:\psi^{\beta_3}\psi^{\mu_3}:\psi^{\beta_6}\psi^{\mu_6}:>g_6^{\mu_4\mu_5}\nonumber\\
(b_{7})^{\mu_3\mu_4\mu_5\mu_6}_{AB}&\!\!\!\!=\!\!\!\!&(ip_2\inn D)_{\beta_4}(ip_3\inn D)_{\beta_6}<:S_A:S_B:\psi^{\beta_4}\psi^{\mu_4}:\psi^{\beta_6}\psi^{\mu_6}:>g_{7}^{\mu_3\mu_5}\nonumber\\
(b_{8})^{\mu_3\mu_4\mu_5\mu_6}_{AB}&\!\!\!\!=\!\!\!\!&2(ip_2)_{\beta_3}(ip_2\inn D)_{\beta_4}(ip_3)_{\beta_5}<:S_A:S_B:\psi^{\beta_3}\psi^{\mu_3}:\psi^{\beta_4}\psi^{\mu_4}:\psi^{\beta_5}\psi^{\mu_5}:>g_{8}^{\mu_6}\nonumber\\
(b_{9})^{\mu_3\mu_4\mu_5\mu_6}_{AB}&\!\!\!\!=\!\!\!\!&2(ip_2)_{\beta_3}(ip_2\inn D)_{\beta_4}(ip_3\inn D)_{\beta_6}<:S_A:S_B:\psi^{\beta_3}\psi^{\mu_3}:\psi^{\beta_4}\psi^{\mu_4}:\psi^{\beta_6}\psi^{\mu_6}:>g_{9}^{\mu_5}\nonumber\\
(b_{10})^{\mu_3\mu_4\mu_5\mu_6}_{AB}&\!\!\!\!=\!\!\!\!&(ip_2)_{\beta_3}(ip_2\inn D)_{\beta_4}(ip_3)_{\beta_5}(ip_3\inn D)_{\beta_6}\nonumber\\
&&\qquad\qquad\qquad\times<:S_A:S_B:\psi^{\beta_3}\psi^{\mu_3}:\psi^{\beta_4}\psi^{\mu_4}:\psi^{\beta_5}\psi^{\mu_5}:\psi^{\beta_6}\psi^{\mu_6}:>g_{10}\nonumber
\eeqa
where $g$'s are  the correlators  of $X$'s which can easily be performed using the standard world-sheet propagators, and  
the correlator of $\psi$ can be calculated using the  Wick-like rule \cite{Liu:2001qa,Garousi:2010bm}.

 Combining the gamma matrices coming from the  Wick-like rule with the gamma matrices in \reef{A}, one finds  the following  trace \cite{Garousi:2010bm}:
 \beqa
T(n,p,m)& =&(H_{1(n)}M_p)^{AB}(\gamma^{\alpha_1\cdots \alpha_m}C^{-1})_{AB}A_{[\alpha_1\cdots \alpha_m]}\labell{relation1}\\
& =&\frac{1}{n!(p+1)!}\veps_{1\nu_1\cdots \nu_{n}}\eps_{a_0\cdots a_p}A_{[\alpha_1\cdots \alpha_m]}\Tr(\gamma^{\nu_1}\cdots \gamma^{\nu_{n}}\gamma^{a_0}\cdots\gamma^{a_p}\gamma^{\alpha_1\cdots \alpha_m})\nonumber
 \eeqa
where $A_{[\alpha_1\cdots \alpha_m]}$ is an antisymmetric combination of the momenta and/or the polarizations of the NSNS states. 
The trace \reef{relation1} can be evaluated for specific values of $n$. One can  verify that the amplitude is non-zero only  for $n=p-3,\, n=p-1,\, n=p+1,\, n=p+3,\, n=p+5$. We are interested in the case
\beqa
n&=&p-3\labell{np3}
\eeqa
The case $n=p+5$ will be  studied in the appendix B.  In above case, the trace relation \reef{relation1} gives non-zero result only for $m\geq 4$.  One immediately concludes that  $b_1,\, b_2$ and $b_3$ in \reef{bs} have no contribution to the amplitude.
The cases that the RR field carries transverse indices are studied in \cite{Garousi:2010bm}. We consider here  the case that the RR potential carries only world volume indices. The gamma matrices in \reef{relation1} corresponding to $\veps_1^{(p-3)}$ must be contracted with the gamma matrices corresponding to the world volume form, otherwise they both   would contract with the gamma matrices  corresponding to $A_{[\alpha_1\cdots \alpha_m]}$ which gives zero result because of the appearance of repeated world volume indices in $A_{[\alpha_1\cdots \alpha_m]}$. In the following we consider the RR scalar field. The result can easily be extended to the RR  $n$-form by contracting its indices with the world volume form.

For the RR scalar $n=0$, and from the relation \reef{np3} one gets $p=3$. The trace  \reef{relation1} is non-zero only for $m=4$. It becomes 
\beqa
T(0,3,4)&=&32\eps^{a_0\cdots a_5}A_{[a_0\cdots a_3]}
\eeqa
where 32 is the trace of the $32\times 32$ identity matrix. Since all indices of $A_{[a_0\cdots a_3]}$ are world volume, one finds $b_{4}$ has no contribution to the amplitude.  The $\psi$ correlators in $b_{10},\,b_{9},\, b_{8},\,b_{7},\, b_{6},\,b_{5}$ have non-zero contributions to the amplitude \reef{amp2}. The $X$ correlator in  $b_{10}$ is 
\beqa
g_{10}&\!\!\!\!=\!\!\!\!&|z_{12}|^{2p_1\cdot p_2}|z_{13}|^{2p_1\cdot p_3}|z_{23}|^{2p_2\cdot p_3}|z_{1\bar2}|^{2p_1\cdot D\cdot p_2}|z_{1\bar3}|^{2p_1\cdot D\cdot p_3}|z_{2\bar3}|^{2p_2\cdot D\cdot p_3} \nonumber\\
&&\times(z_{1\bar1})^{p_1\cdot D\cdot p_1}(z_{2\bar2})^{p_2\cdot D\cdot p_2}(z_{3\bar3})^{p_3\cdot D\cdot p_3}(i)^{p_1\cdot D\cdot p_1+p_2\cdot D\cdot p_2+p_3\cdot D\cdot p_3}\equiv K\labell{K}
\eeqa
where $z_{ij}=z_i-z_j$ and $z_{i\bar j}=z_i-\bz_j$. We have added the phase factor to make it real. The above function  appears in all other $X$ correlators in \reef{bs}. 
The  $\psi$ correlators in $b_{10}$ gives $24\times 12$ terms which result from different Wick-like contractions. The contractions which end up with having $p_2$ and $ p_2\inn D$ or $p_3$ and $ p_3\inn D$ in $A_{[a_0\cdots a_3]}$ give zero result.

The $X$ correlators in $b_{8},\, b_{9}$ are \cite{Garousi:2010bm}
\beqa
g_8^{\mu_6}&=&  \frac{iK}{z_{3\bar3}}\left(\frac{p_1^{\mu _6}z_{31}}{z_{1\bar3}}+\frac{p_2^{\mu _6}z_{32}}{z_{2\bar3}}+\frac{(p_1\inn D)^{\mu _6}z_{3\bar1}}{{z}_{\bar1\bar3}}+\frac{(p_2\inn D)^{\mu _6}z_{3\bar2}}{{z}_{\bar2\bar3}}\right)\nonumber\\
g_9^{\mu_5}&=&  \frac{iK}{z_{\bar3 3}}\left(\frac{p_1^{\mu _5}z_{\bar3 1}}{z_{13}}+\frac{p_2^{\mu _5}z_{\bar3 2}}{z_{23}}+\frac{(p_1\inn D)^{\mu _5}z_{\bar3\bar1}}{{z}_{\bar1 3}}+\frac{(p_2\inn D)^{\mu _5}z_{\bar3\bar2}}{{z}_{\bar2 3}}\right)\nonumber
\eeqa
The  $\psi$ correlators in each of $b_{8},\, b_{9}$ gives $ 12$ terms which result from different Wick-like contractions. 
 
The $X$ correlators in $b_{5},\, b_{6},\,b_{7}$ are 
 \beqa
g_{5}^{\mu_4\mu_6}&=&-\frac{\eta^{\mu_4\mu_6}K}{z_{\bar2\bar3}^2}-\frac{K}{z_{2\bar2}z_{3\bar3}}\left(\frac{p_1^{\mu_4}z_{21}}{z_{1\bar2}}+\frac{p_3^{\mu_4}z_{23}}{z_{3\bar2}}+\frac{(p_1\inn D)^{\mu_4}z_{2\bar1}}{z_{\bar1\bar2}}+\frac{(p_3\inn D)^{\mu_4}z_{2\bar3}}{z_{\bar3\bar2}}\right)\nonumber\\
 &&\qquad\qquad\qquad\quad\times\left(\frac{p_1^{\mu _6}z_{31}}{z_{1\bar3}}+\frac{p_2^{\mu _6}z_{32}}{z_{2\bar3}}+\frac{(p_1\inn D)^{\mu_6}z_{3\bar1}}{{z}_{\bar1\bar3}}+\frac{(p_2\inn D)^{\mu _6}z_{3\bar2}}{{z}_{\bar2\bar3}}\right)\nonumber\\
g_{6}^{\mu_4\mu_5}&=&-\frac{\eta^{\mu_4\mu_5}K}{z_{\bar2 3}^2}-\frac{K}{z_{2\bar2}z_{\bar3 3}}\left(\frac{p_1^{\mu_4}z_{21}}{z_{1\bar2}}+\frac{p_3^{\mu_4}z_{23}}{z_{3\bar2}}+\frac{(p_1\inn D)^{\mu_4}z_{2\bar1}}{z_{\bar1\bar2}}+\frac{(p_3\inn D)^{\mu_4}z_{2\bar3}}{z_{\bar3\bar2}}\right)\nonumber\\
 &&\qquad\qquad\qquad\quad\times\left(\frac{p_1^{\mu _5}z_{\bar3 1}}{z_{13}}+\frac{p_2^{\mu _5}z_{\bar3 2}}{z_{23}}+\frac{(p_1\inn D)^{\mu _5}z_{\bar3\bar1}}{{z}_{\bar1 3}}+\frac{(p_2\inn D)^{\mu _5}z_{\bar3\bar2}}{{z}_{\bar2 3}}\right)\nonumber\\ 
 g_{7}^{\mu_3\mu_5}&=&-\frac{\eta^{\mu_3\mu_5}K}{z_{2 3}^2}-\frac{K}{z_{\bar2 2}z_{\bar3 3}}\left(\frac{p_1^{\mu_3}z_{\bar2 1}}{z_{12}}+\frac{p_3^{\mu_3}z_{\bar2 3}}{z_{32}}+\frac{(p_1\inn D)^{\mu_3}z_{\bar2\bar1}}{z_{\bar1 2}}+\frac{(p_3\inn D)^{\mu_3}z_{\bar2\bar3}}{z_{\bar3 2}}\right)\nonumber\\
 &&\qquad\qquad\qquad\quad\times\left(\frac{p_1^{\mu _5}z_{\bar3 1}}{z_{13}}+\frac{p_2^{\mu _5}z_{\bar3 2}}{z_{23}}+\frac{(p_1\inn D)^{\mu _5}z_{\bar3\bar1}}{{z}_{\bar1 3}}+\frac{(p_2\inn D)^{\mu _5}z_{\bar3\bar2}}{{z}_{\bar2 3}}\right)\labell{g567}
 \eeqa
The  $\psi$ correlators in each of them  gives one term. Examining  the transformation of the above X-correlators and the correlators of $\psi$'s in Wick-like rule, one can easily verify that the amplitude $\reef{A}$ is invariant under the $SL(2,R)$ transformation.  So one can map the results to disk with unit radius. That is,  one can use the following replacement\cite{Garousi:2010bm}:
\beqa
z_{i\bar j}&\rightarrow &-(1-z_i\bz_j)\nonumber\\
z_{ij}&\rightarrow &z_{ij}\nonumber\\
z_{\bar i\bar j}&\rightarrow &-z_{\bar i\bar j}\nonumber\\
z_{\bar i  j}&\rightarrow &(1-\bz_iz_j)\nonumber
\eeqa
Obviously the result is still  $SL(2,R)$ invariant. To fix this symmetry, we then set \cite{Craps:1998fn}
\beqa
z_1=0,&{\rm and}& z_2=\bz_2=r_2\labell{fix}
\eeqa
Under this fixing the measure in \reef{A} changes as
\beqa
d^2z_1d^2z_2d^2z_3\rightarrow r_2dr_2\,r_3dr_3\,d\theta,&&0<r_2,r_3<1,\,0<\theta<2\pi
\eeqa
where we have chosen  the polar coordinate $z_3=r_3e^{i\theta}$, and  $K$ changes as  
\beqa
\tK&\!\!\!\!=\!\!\!\!&\ {r_2}^{2p_1\cdot p_2}\ {r_3}^{2 p_1\cdot p_3} (1-{r_2}^2)^{p_2\cdot D\cdot p_2}(1-{r_3}^2)^{p_3\cdot D\cdot p_3}\nonumber\\
&&\left.\times |r_2-r_3 e^{i \t}|^{2 p_2\cdot p_3}|1-r_2 r_3 e^{i \t}|^{2 p_2\cdot D\cdot p_3}\right. \labell{KK}
\eeqa

The first terms in \reef{g567}  produce structures in which the  polarization tensors contract with each other. Let us consider these terms. They appear in the amplitude as
\beqa
{\cal A}_{5}(\veps_2\inn\veps_3^T)&\sim&-4\eps_{a_0\cdots a_3}p_2^{a_0}p_3^{a_1}(\veps_2\inn\veps_3^T)^{a_2a_3}\int d^2z_1d^2z_2d^2z_3\frac{K}{z_{\bar2\bar3}^2z_{21}z_{2\bar1}z_{31}z_{3\bar1}}\nonumber\\
{\cal A}_{6}(\veps_2\inn D\inn\veps_3)&\sim&-8\eps_{a_0\cdots a_3}p_2^{a_0}p_3^{a_1}(\veps_2\inn D\inn\veps_3)^{a_2a_3}\int d^2z_1d^2z_2d^2z_3\frac{K}{z_{\bar2 3}^2z_{21}z_{2\bar1}z_{\bar3 1}z_{\bar3\bar1}}\nonumber\\
{\cal A}_{7}(\veps_2^T\inn\veps_3)&\sim&-4\eps_{a_0\cdots a_3}p_2^{a_0}p_3^{a_1}(\veps_2^T\inn\veps_3)^{a_2a_3}\int d^2z_1d^2z_2d^2z_3\frac{K}{z_{23}^2z_{\bar2 1}z_{\bar2\bar1}z_{\bar3 1}z_{\bar3\bar1}}\labell{A567}
\eeqa
where we have written the sub-amplitudes corresponding to $b_i$ in \reef{A} as ${\cal A}_i$. The above sub-amplitudes  are zero when one polarization is symmetric and the other one is antisymmetric. 

There are other sub-amplitudes which   have  also terms in which the polarization tensor contract with each other. 
The  amplitudes \reef{A567} have second order poles, \eg $1/(z_{\bar2\bar3})^2$ in  ${\cal A}_5$. They indicates that there is a tachyon propagating in the amplitude. This undesirable feather appears when one uses the vertex operator in 0-picture \cite{Green:1987qu}. However, the whole amplitude \reef{A} has no tachyon which means all the tachyons in the sub-amplitudes must  be canceled among them. So one may  keep the second order poles in the sub-amplitudes and  the tachyons would be canceled  finally  using the  properties of the functions that appear in the final amplitude, \eg in the four point function one has $(p_2\inn p_3-1)\Ga(p_2\inn p_3-1)=\Ga(p_2\inn p_3)$ where $\Ga(p_2\inn p_3-1)$ has tachyon pole whereas $\Ga(p_2\inn p_3)$ has massless pole. 

Alternatively, one can show that the second order poles appear in the whole amplitude as derivative of first order poles. Then using by part integration, one can write them in terms of first order poles. In this case, one needs to use a by part integration to remove the tachyon.  Mapping the above amplitudes  to unit disk and fixing the $SL(2,R)$ symmetry as \reef{fix}, one can write \reef{A567}, after a by part integration, as  
\beqa
{\cal A}_5(\veps_2\inn\veps_3^T)&\sim&- 4\eps_{a_0\cdots a_3}p_2^{a_0}p_3^{a_1}(\veps_2\inn\veps_3^T)^{a_2a_3}\int dr_2dr_3 d\theta\frac{\frac{\prt\tK}{\prt i\theta}}{r_3(r_2-r_3e^{-i\theta})} \nonumber\\
{\cal A}_6(\veps_2\inn D\inn\veps_3)&\sim&8\eps_{a_0\cdots a_3}p_2^{a_0}p_3^{a_1}(\veps_2\inn D\inn\veps_3)^{a_2a_3}\int dr_2dr_3 d\theta\frac{\frac{\prt\tK}{\prt i\theta}}{r_3(r_2-r_3e^{i\theta})} \nonumber\\
{\cal A}_7(\veps_2^T\inn\veps_3)&\sim& 4\eps_{a_0\cdots a_3}p_2^{a_0}p_3^{a_1}(\veps_2^T\inn\veps_3)^{a_2a_3}\int dr_2dr_3 d\theta\frac{\frac{\prt\tK}{\prt i\theta}}{r_3(r_2-r_3e^{i\theta})}\labell{A567b}
\eeqa 
where 
\beqa
\frac{\prt\tK}{\prt i\theta}&=&r_2r_3(e^{i\theta}-e^{-i\theta})\left(\frac{p_2\inn p_3}{|r_2-r_3e^{i\theta}|^2}+\frac{p_2\inn D\inn p_3}{|1-r_2r_3e^{i\theta}|^2}\right)\tK
\eeqa
Note that the  sub-amplitudes \reef{A567} have two momenta whereas the sub-amplitudes \reef{A567b} have four momenta, as all the other structures in the amplitude \reef{A}.

Since there is no conservation of momentum in the transverse directions in \reef{A}, the terms in  which $p_1^i$ contracts with each of the polarizations, \ie  $p_1\inn N\inn\veps_3,\, p_1\inn N\inn\veps_2$, as well as  $p_2\inn N\inn\veps_3,\, p_3\inn N\inn\veps_2$ are independent structures. For the other terms we use  the conservation of momentum along the brane, \ie
\beqa
(p_1+p_2+p_3)\inn V_{\mu}&=&0\labell{cons}
\eeqa
 to write $p_1^a$ in terms of $p_2^a$ and $p_3^a$. 
 
 When one tensor is symmetric and the other one is antisymmetric the result is zero. The result for two symmetric tensors is
 \beqa
 A&\sim&\eps_{a_0a_1a_2a_3}p_2^{a_0}p_3^{a_1}\bigg[p_2\inn N\inn p_3(\veps_2\inn N\inn\veps_3)^{a_2a_3}+p_2\inn V\inn p_3(\veps_2\inn V\inn\veps_3)^{a_2a_3}\nonumber\\
 &&-p_3\inn N\inn\veps_2^{a_2}\,p_2\inn N\inn\veps_3^{a_3}-p_3\inn V\inn\veps_2^{a_2}\,p_2\inn V\inn\veps_3^{a_3}\bigg]\cJ\labell{gg}
 \eeqa
 where $\cJ$ is 
 \beqa
 \cJ&=&-4\int_0^1dr_2\int_0^{1}dr_3\,r_2r_3\int_0^{2\pi}d\theta\frac{\sin^2(\theta)\tK}{|1-r_2r_3e^{i\theta}|^2|r_2-r_3e^{-i\theta}|^2}\nonumber
 \eeqa
 This is the result that has been found in  \cite{Craps:1998fn}. It is shown in \cite{Craps:1998fn} that the above integral  has only contact term at low energy, \ie
\beqa
\cJ&=&-\frac{\pi^3}{3}+\cdots\labell{exp12}
\eeqa 
where dots represent terms with two and more momenta which correspond to the amplitude at order $O(\alpha'^3)$ in which we are not interested. It has been shown in \cite{Craps:1998fn} that the above contact terms reproduce the gravity couplings in \reef{Tf2}.

The indices of the graviton polarization tensors in the second line of \reef{gg} are contracted with the world volume form or with the momentum. This indicates that these terms are invariant under linear T-duality when the Killing coordinate is an index of the RR potential \cite{Garousi:2010rn}. When the Killing coordinate is an index of the graviton polarization tensor, T-duality relates them to the higher RR form \cite{Garousi:2010rn} in which we are not interested in this paper. On the other hand, one of the  indices of the graviton polarization tensors in the first line of \reef{gg} are contracted with each other. This indicates that the terms in the first line are not invariant under T-duality \cite{Garousi:2010rn}. Hence, there must be antisymmetric tensor  couplings as well to make them invariant. 

However, the amplitude  for two antisymmetric tensors has much more terms than those that are needed to make the gravity couplings to be invariant under linear T-duality. The result  is
\beqa
{\cal A}&\!\!\!\!\!\sim\!\!\!\!\!&\frac{1}{2}\eps_{a_0a_1a_2a_3}\veps_3^{a_2a_3}\bigg(p_2^{a_0}p_3^{a_1}p_3\inn V\inn\veps_2\inn V\inn p_2\cJ_1-\frac{1}{2}p_2^{a_0}p_3^{a_1}p_3\inn V\inn\veps_2\inn N\inn p_1\cI_3\labell{total}\\
&\!\!\!\!\!\!\!\!\!\!&+p_2^{a_0}p_3^{a_1}p_2\inn V\inn\veps_2\inn N\inn p_3\cJ_2-p_2^{a_0}p_3^{a_1}p_3\inn V\inn\veps_2\inn N\inn p_3\cJ_0+\frac{1}{2}p_2^{a_0}p_3^{a_1}p_3\inn N\inn\veps_2\inn N\inn p_1\cI_2\nonumber\\
&\!\!\!\!\!\!\!\!\!\!&-2p_2^{a_0}p_3\inn V\inn p_3\,p_2\inn V\inn\veps_2^{a_1}\cJ_3+\frac{1}{2}p_2^{a_0}p_3\inn V\inn p_3\,p_3\inn V\inn\veps_2^{a_1}\cJ_4+\frac{1}{2}p_3^{a_0}p_2\inn V\inn p_2\,p_3\inn V\inn\veps_2^{a_1}\cJ_1\nonumber\\
&\!\!\!\!\!\!\!\!\!\!&+p_2^{a_0}p_3\inn V\inn p_3\,p_1\inn N\inn\veps_2^{a_1}\cI_4-\frac{1}{2}p_2^{a_0}p_3\inn V\inn p_3\,p_3\inn N\inn\veps_2^{a_1}\cJ_{12}-\frac{1}{2}p_3^{a_0}p_2\inn V\inn p_2\,p_3\inn N\inn\veps_2^{a_1}\cJ_2\nonumber\\
&\!\!\!\!\!\!\!\!\!\!&+\frac{1}{2}p_3^{a_0}p_2\inn V\inn p_3\,p_1\inn N\inn\veps_2^{a_1}\cI_3-\frac{1}{2}p_3^{a_0}p_2\inn N\inn p_3\,p_1\inn N\inn\veps_2^{a_1}\cI_2-p_3^{a_0}p_2\inn V\inn p_3\,p_2\inn V\inn\veps_2^{a_1}\cJ_1\nonumber\\
&\!\!\!\!\!\!\!\!\!\!&+p_3^{a_0}p_2\inn N\inn p_3\,p_2\inn V\inn\veps_2^{a_1}\cJ_2-\frac{1}{2}p_2^{a_0}p_2\inn V\inn p_3\,p_3\inn V\inn\veps_2^{a_1}\cJ_6-\frac{1}{2}p_3^{a_0}p_2\inn V\inn p_3\,p_3\inn V\inn\veps_2^{a_1}\cJ_6\nonumber\\
&\!\!\!\!\!\!\!\!\!\!&-p_2^{a_0}p_2\inn N\inn p_3\,p_3\inn V\inn\veps_2^{a_1}\cJ_7-p_3^{a_0}p_2\inn N\inn p_3\,p_3\inn V\inn\veps_2^{a_1}\cJ_8-p_2^{a_0}p_2\inn V\inn p_3\,p_3\inn N\inn\veps_2^{a_1}\cJ_9\nonumber\\
&\!\!\!\!\!\!\!\!\!\!&-p_3^{a_0}p_2\inn V\inn p_3\,p_3\inn N\inn\veps_2^{a_1}\cJ_{10}-\frac{1}{2}p_2^{a_0}p_2\inn N\inn p_3\,p_3\inn N\inn\veps_2^{a_1}\cJ_6-\frac{1}{2}p_3^{a_0}p_2\inn N\inn p_3\,p_3\inn N\inn\veps_2^{a_1}\cJ_6\nonumber\\
&\!\!\!\!\!\!\!\!\!\!&+\frac{1}{8}(p_2\inn V\inn p_3)^2\,\veps_2^{a_0a_1}\cJ_6+\frac{1}{8}(p_2\inn N\inn p_3)^2\,\veps_2^{a_0a_1}\cJ_6+\frac{1}{4}p_2\inn N\inn p_3\,p_2\inn V\inn p_3\,\veps_2^{a_0a_1}\cJ_{11}\nonumber\\
&\!\!\!\!\!\!\!\!\!\!&+\frac{1}{4}p_2\inn V\inn p_2\,p_3\inn V\inn p_3\,\veps_2^{a_0a_1}\cJ_3\bigg)\nonumber\\
&\!\!\!\!\!\!\!\!\!\!&+\frac{1}{2}\eps_{a_0a_1a_2a_3}p_2^{a_0}p_3^{a_1}\bigg(-p_1\inn N\inn\veps_2^{a_2}\,p_2\inn V\inn\veps_3^{a_3}\cI_3-p_1\inn N\inn\veps_2^{a_2}\,p_1\inn N\inn\veps_3^{a_3}\cI_1\nonumber\\
&\!\!\!\!\!\!\!\!\!\!&+p_1\inn N\inn\veps_2^{a_2}\,p_2\inn N\inn\veps_3^{a_3}\cI_2+2p_3\inn V\inn\veps_2^{a_2}\,p_2\inn N\inn\veps_3^{a_3}\cJ_5+4p_1\inn N\inn\veps_2^{a_2}\,p_3\inn V\inn\veps_3^{a_3}\cI_4\nonumber\\
&\!\!\!\!\!\!\!\!\!\!&-2p_2\inn V\inn\veps_2^{a_2}\,p_2\inn N\inn\veps_3^{a_3}\cJ_2+2p_2\inn V\inn\veps_2^{a_2}\,p_2\inn V\inn\veps_3^{a_3}\cJ_1-4p_2\inn V\inn\veps_2^{a_2}\,p_3\inn V\inn\veps_3^{a_3}\cJ_3\nonumber\\
&\!\!\!\!\!\!\!\!\!\!&+p_2\inn N\inn p_3(\veps_2\inn V\inn\veps_3)^{a_2a_3}\cJ-p_2\inn V\inn p_3(\veps_2\inn N\inn\veps_3)^{a_2a_3}\cJ\bigg)+(2\leftrightarrow 3)\nonumber
\eeqa 
The indices of the  polarization tensors in all terms except the terms in the last line are contracted with the world volume form or with the momentum. Hence they all are invariant under linear T-duality. The terms in the last line combines with the terms in the first line of \reef{gg} to make a T-dual combination \cite{Becker:2010ij,Garousi:2010rn}. However, the above amplitude has more couplings than those have been found in  \cite{Becker:2010ij,Garousi:2010rn} by requiring the consistency of the graviton couplings in the Chern-Simons action with linear T-duality. The new  couplings  which are invariant under linear T-duality can be found by studying the  integrals that appear in the amplitude.

The integral $\cJ$ is the one which appears also in the graviton amplitude, and the integrals $\cI_1,\cdots ,\cI_4$ in \reef{total} are those  which appear also in the scattering amplitude considered in  \cite{Garousi:2010bm} in which the RR potential carries both transverse and world volume indices. These integrals are 
\beqa
\cI_1&=&-\int_0^1dr_2\int_0^1dr_3\frac{1}{r_2r_3}\int_0^{2\pi}d\theta\tK\nonumber\\
\cI_2&=&2\int_0^1dr_2\int_0^{1}dr_3\frac{(1-r_2^2)}{r_2}\int_0^{2\pi}d\theta\frac{[r_3(1+r_2^2)-r_2(1+r_3^2)\cos(\theta)]\tK}{|1-r_2r_3e^{i\theta}|^2|r_2-r_3e^{-i\theta}|^2}\nonumber\\
\cI_4&=&-\int_0^1dr_2\int_0^{1}dr_3\frac{(1+r_3^2)}{r_2r_3(1-r_3^2)}\int_0^{2\pi}d\theta\tK\nonumber
\eeqa
and $\cI_3(p_1,p_2,p_3)=\cI_2(p_1,p_3,p_2)$. The other integrals are
\beqa
\cJ_0&=&\int_0^1dr_2\int_0^{1}dr_3\frac{1}{r_2r_3}\int_0^{2\pi}d\theta\frac{[(1-r_2^2r_3^2)(r_2^2-r_3^2)-4r_2^2r_3^2\sin^2(\theta)]\tK}{|1-r_2r_3e^{i\theta}|^2|r_2-r_3e^{-i\theta}|^2}\nonumber\\
\cJ_1&=&2\int_0^1dr_2\int_0^{1}dr_3\frac{(1+r_2^2)(1-r_3^2)}{r_3(1-r_2^2)}\int_0^{2\pi}d\theta\frac{[r_2(1+r_3^2)-r_3(1+r_2^2)\cos(\theta)]\tK}{|1-r_2r_3e^{i\theta}|^2|r_2-r_3e^{-i\theta}|^2}\nonumber\\
\cJ_2&=&2\int_0^1dr_2\int_0^{1}dr_3\frac{(1+r_2^2)}{r_2}\int_0^{2\pi}d\theta\frac{[r_3(1+r_2^2)-r_2(1+r_3^2)\cos(\theta)]\tK}{|1-r_2r_3e^{i\theta}|^2|r_2-r_3e^{-i\theta}|^2}\nonumber\\
\cJ_3&=&-\int_0^1dr_2\int_0^{1}dr_3\frac{(1+r_3^2)(1+r_2^2)}{r_2r_3(1-r_3^2)(1-r_2^2)}\int_0^{2\pi}d\theta\tK\nonumber\\
\cJ_5&=&\int_0^1dr_2\int_0^{1}dr_3\frac{(1-r_2^2r_3^2)(r_2^2-r_3^2)}{r_2r_3}\int_0^{2\pi}d\theta\frac{\tK}{|1-r_2r_3e^{i\theta}|^2|r_2-r_3e^{-i\theta}|^2}\nonumber\\
\cJ_6&=&-2\int_0^1dr_2\int_0^{1}dr_3(1-r_2^2)(1-r_3^2)\int_0^{2\pi}d\theta\frac{\cos(\theta)\tK}{|1-r_2r_3e^{i\theta}|^2|r_2-r_3e^{-i\theta}|^2}\nonumber\\
\cJ_7&=&-\int_0^1dr_2\int_0^{1}dr_3\frac{r_2}{r_3}\int_0^{2\pi}d\theta\frac{[(1-r_3^2)^2+4r_3^2\sin^2(\theta)]\tK}{|1-r_2r_3e^{i\theta}|^2|r_2-r_3e^{-i\theta}|^2}\nonumber\\
\cJ_9&=&-\int_0^1dr_2\int_0^{1}dr_3\frac{(1-r_2^2)^2r_3}{r_2}\int_0^{2\pi}d\theta\frac{\tK}{|1-r_2r_3e^{i\theta}|^2|r_2-r_3e^{-i\theta}|^2}\nonumber\\
\cJ_{11}&=&-\int_0^1dr_2\int_0^{1}dr_3\frac{1}{r_2r_3}\int_0^{2\pi}d\theta\frac{[(1+r_2^2r_3^2)(r_2^2+r_3^2)-4r_2^2r_3^2+4r_2^2r_3^2\sin^2(\theta)]\tK}{|1-r_2r_3e^{i\theta}|^2|r_2-r_3e^{-i\theta}|^2}\nonumber
\eeqa
and
\beqa
&&\cJ_4(p_1,p_2,p_3)=\cJ_1(p_1,p_3,p_2)\,;\,\cJ_{12}(p_1,p_2,p_3)=\cJ_2(p_1,p_3,p_2)\nonumber\\
&&\cJ_8(p_1,p_2,p_3)=\cJ_7(p_1,p_3,p_2)\,;\,\cJ_{10}(p_1,p_2,p_3)=\cJ_9(p_1,p_3,p_2)
\eeqa
It is shown in \cite{Garousi:2010bm} that the integrals $\cI_1, \cdots, \cI_4$ have no contact terms. However, as we will see later  some of the integrals $\cJ_0,\cdots, \cJ_{11}$ have contact terms.

The amplitude \reef{total} should satisfy the Ward identities associated with the RR field and with the B-fields. If one could perform the integrals explicitly, then one would be able to check these identities explicitly. Alternatively, by demanding the amplitude \reef{total} to satisfy these Ward identities, one would be able to find  some relations between the integrals. Checking these relations explicitly would confirm the amplitude satisfies the Ward identities. We use the latter method in this paper. 

The relations between the integrals may be used to write the amplitude \reef{total} either in terms of RR field strength, $F$, or in terms of field strength of the B-field, $H$. Since the relations between the integrals involve only Mandelstam variables $p_2\inn V\inn p_2$, $ p_3\inn V\inn p_3$, $\cdots$, we expect the terms in the amplitude \reef{total} which have no Mandelstam variables can easily be written in terms of $H$. For example, the first term in \reef{total} can be written as $H_3^{a_1a_2a_3}H_2^{a_0ab}p_{2a}p_{3b}/3$. This term includes the first term in \reef{total} and some extra terms which are proportional to the Mandelstam variables. The contribution of all such terms should be canceled in the amplitude after using the relation between the integrals. We will see that in this way one is able to write the amplitude in terms of $H$.

Before finding the relations between the integrals, we reduce the number of integrals involved in the amplitude \reef{total}. One observes that the integrals $\cJ_0,\,\cJ_7,\,\cJ_8,\,\cJ_{11}$  include $\sin(\theta)^2$. This part of the integrals is exactly  $\cJ$. Separating this part, one can  rewrite the amplitude in the following form:
\beqa
{\cal A}&\!\!\!\!\!\sim\!\!\!\!\!&\frac{1}{2}\eps_{a_0a_1a_2a_3}\bigg[\veps_3^{a_2a_3}\bigg(-p_2^{a_0}p_3^{a_1}p_3\inn V\inn\veps_2\inn N\inn p_3-p_2^{a_0}p_2\inn N\inn p_3\,p_3\inn V\inn\veps_2^{a_1}\nonumber\\
&\!\!\!\!\!\!\!\!\!\!&-p_3^{a_0}p_2\inn N\inn p_3\,p_3\inn V\inn\veps_2^{a_1}
+\frac{1}{4}p_2\inn N\inn p_3\,p_2\inn V\inn p_3\,\veps_2^{a_0a_1}\bigg)\nonumber\\
&\!\!\!\!\!\!\!\!\!\!&+p_2^{a_0}p_3^{a_1}\bigg(
p_2\inn N\inn p_3(\veps_2\inn V\inn\veps_3)^{a_2a_3}-p_2\inn V\inn p_3(\veps_2\inn N\inn\veps_3)^{a_2a_3}\bigg)\bigg]\cJ\nonumber\\
&\!\!\!\!\!\!\!\!\!\!&+\frac{1}{2}\eps_{a_0a_1a_2a_3}\veps_3^{a_2a_3}\bigg[p_2^{a_0}p_3^{a_1}p_3\inn V\inn\veps_2\inn V\inn p_2\cJ_1-\frac{1}{2}p_2^{a_0}p_3^{a_1}p_3\inn V\inn\veps_2\inn N\inn p_1\cI_3\labell{totalpole}\\
&\!\!\!\!\!\!\!\!\!\!&+p_2^{a_0}p_3^{a_1}p_2\inn V\inn\veps_2\inn N\inn p_3\cJ_2-p_2^{a_0}p_3^{a_1}p_3\inn V\inn\veps_2\inn N\inn p_3\cJ_5+\frac{1}{2}p_2^{a_0}p_3^{a_1}p_3\inn N\inn\veps_2\inn N\inn p_1\cI_2\nonumber\\
&\!\!\!\!\!\!\!\!\!\!&-2p_2^{a_0}p_3\inn V\inn p_3\,p_2\inn V\inn\veps_2^{a_1}\cJ_3+\frac{1}{2}p_2^{a_0}p_3\inn V\inn p_3\,p_3\inn V\inn\veps_2^{a_1}\cJ_4+\frac{1}{2}p_3^{a_0}p_2\inn V\inn p_2\,p_3\inn V\inn\veps_2^{a_1}\cJ_1\nonumber\\
&\!\!\!\!\!\!\!\!\!\!&+p_2^{a_0}p_3\inn V\inn p_3\,p_1\inn N\inn\veps_2^{a_1}\cI_4-\frac{1}{2}p_2^{a_0}p_3\inn V\inn p_3\,p_3\inn N\inn\veps_2^{a_1}\cJ_{12}-\frac{1}{2}p_3^{a_0}p_2\inn V\inn p_2\,p_3\inn N\inn\veps_2^{a_1}\cJ_2\nonumber\\
&\!\!\!\!\!\!\!\!\!\!&+\frac{1}{2}p_3^{a_0}p_2\inn V\inn p_3\,p_1\inn N\inn\veps_2^{a_1}\cI_3-\frac{1}{2}p_3^{a_0}p_2\inn N\inn p_3\,p_1\inn N\inn\veps_2^{a_1}\cI_2-p_3^{a_0}p_2\inn V\inn p_3\,p_2\inn V\inn\veps_2^{a_1}\cJ_1\nonumber\\
&\!\!\!\!\!\!\!\!\!\!&+p_3^{a_0}p_2\inn N\inn p_3\,p_2\inn V\inn\veps_2^{a_1}\cJ_2+\frac{1}{4}p_2^{a_0}p_2\inn  p_3\,p_3\inn D\inn\veps_2^{a_1}\cJ_5-\frac{1}{4}p_3^{a_0}p_2\inn  p_3\,p_3\inn D\inn\veps_2^{a_1}\cJ_{5}\nonumber\\
&\!\!\!\!\!\!\!\!\!\!&+\frac{1}{4}p_2^{a_0}p_2\inn  p_3\,p_3\inn \veps_2^{a_1}\cJ_{13}+\frac{1}{4}p_3^{a_0}p_2\inn  p_3\,p_3\inn \veps_2^{a_1}\cJ_{13}-\frac{1}{4}p_2^{a_0}p_2\inn D\inn p_3\,p_3\inn D\inn\veps_2^{a_1}\cJ_{14}\nonumber\\
&\!\!\!\!\!\!\!\!\!\!&-\frac{1}{4}p_3^{a_0}p_2\inn D\inn p_3\,p_3\inn D\inn\veps_2^{a_1}\cJ_{14}-\frac{1}{4}p_2^{a_0}p_2\inn D\inn p_3\,p_3\inn \veps_2^{a_1}\cJ_5+\frac{1}{4}p_3^{a_0}p_2\inn D\inn p_3\,p_3\inn \veps_2^{a_1}\cJ_{5}\nonumber\\
&\!\!\!\!\!\!\!\!\!\!&-\frac{1}{16}(p_2\inn  p_3)^2\,\veps_2^{a_0a_1}\cJ_{13}+\frac{1}{16}(p_2\inn D\inn p_3)^2\,\veps_2^{a_0a_1}\cJ_{14}+\frac{1}{4}p_2\inn V\inn p_2\,p_3\inn V\inn p_3\,\veps_2^{a_0a_1}\cJ_3\bigg]\nonumber\\
&\!\!\!\!\!\!\!\!\!\!&+\frac{1}{2}\eps_{a_0a_1a_2a_3}p_2^{a_0}p_3^{a_1}\bigg[-p_1\inn N\inn\veps_2^{a_2}\,p_2\inn V\inn\veps_3^{a_3}\cI_3-p_1\inn N\inn\veps_2^{a_2}\,p_1\inn N\inn\veps_3^{a_3}\cI_1\nonumber\\
&\!\!\!\!\!\!\!\!\!\!&+p_1\inn N\inn\veps_2^{a_2}\,p_2\inn N\inn\veps_3^{a_3}\cI_2+2p_3\inn V\inn\veps_2^{a_2}\,p_2\inn N\inn\veps_3^{a_3}\cJ_5+4p_1\inn N\inn\veps_2^{a_2}\,p_3\inn V\inn\veps_3^{a_3}\cI_4\nonumber\\
&\!\!\!\!\!\!\!\!\!\!&-2p_2\inn V\inn\veps_2^{a_2}\,p_2\inn N\inn\veps_3^{a_3}\cJ_2+2p_2\inn V\inn\veps_2^{a_2}\,p_2\inn V\inn\veps_3^{a_3}\cJ_1-4p_2\inn V\inn\veps_2^{a_2}\,p_3\inn V\inn\veps_3^{a_3}\cJ_3\bigg]+(2\leftrightarrow 3)\nonumber
\eeqa 
where
\beqa
\cJ_{15}(p_1,p_2,p_3)=\cJ_5(p_1,p_3,p_2)=-\cJ_5(p_1,p_2,p_3)\nonumber
\eeqa
and
\beqa
\cJ_{13}\!=\!\!\!\int_0^1dr_2\int_0^{1}dr_3\frac{1}{r_2r_3}\int_0^{2\pi}d\theta\frac{[(1+r_2^2r_3^2)(r_2^2+r_3^2)-4r_2^2r_3^2+2r_2r_3(1-r_2^2)(1-r_3^2)\cos(\theta)]\tK}{|1-r_2r_3e^{i\theta}|^2|r_2-r_3e^{-i\theta}|^2}\nonumber\\
\cJ_{14}\!=\!\!\!\int_0^1dr_2\int_0^{1}dr_3\frac{1}{r_2r_3}\int_0^{2\pi}d\theta\frac{[(1+r_2^2r_3^2)(r_2^2+r_3^2)-4r_2^2r_3^2-2r_2r_3(1-r_2^2)(1-r_3^2)\cos(\theta)]\tK}{|1-r_2r_3e^{i\theta}|^2|r_2-r_3e^{-i\theta}|^2}\nonumber
\eeqa
Instead of integrals $\cJ_0,\, \cJ_6,\, \cJ_7,\, \cJ_9,\, \cJ_{11}$ which appear in \reef{total}, the amplitude \reef{totalpole} has the integrals $\cJ_{13},\, \cJ_{14}$. As we will see, these integrals are easier to perform explicitly. 

We now find the relations between the integrals.  The  amplitude \reef{totalpole} must satisfy the Ward identity corresponding to polarizations $\veps_2$ and $\veps_3$. Imposing these conditions, one finds some relations between the integrals. The Ward identity for $\veps_2$ gives the following relations:
\beqa
&&-2p_1\inn N\inn p_2\cI_1+2p_2\inn V\inn p_2\cI_7+p_2\inn N\inn p_3\cI_3-p_2\inn V\inn p_3\cI_2=0\labell{iden3}\\
&&-2 \cI_2 p_1\inn N\inn p_2+(\cJ_{13}-\cJ_{14}) p_2\inn N\inn p_3+2 \cJ_2 p_2\inn V\inn p_2\nonumber\\
&&\qquad\qquad\qquad\qquad+(-4 \cJ+\cJ_{13}+\cJ_{14}-2 \cJ_{5}) p_2\inn V\inn p_3=0\nonumber\\
&&2 \cI_{3} p_1\inn N\inn p_2-2 \cJ_{1} p_2\inn V\inn p_2+(\cJ_{13}-\cJ_{14}) p_2\inn V\inn p_3\nonumber\\
&&\qquad\qquad\qquad\qquad+(\cJ_{13}+\cJ_{14}+2 \cJ_{5}) p_2\inn N\inn p_3=0\nonumber\\
&&-2 \cI_{4} p_1\inn N\inn p_2+\cJ_{12} p_2\inn N\inn p_3+2 \cJ_{3} p_2\inn V\inn p_2-\cJ_{4} p_2\inn V\inn p_3=0\nonumber\\
&&(-\cJ_{13}+\cJ_{14}) \left((p_2\inn N\inn p_3)^2+(p_2\inn V\inn p_3)^2\right)+2 p_1\inn N\inn p_2 \left(\cI_{2} p_2\inn N\inn p_3-\cI_{3} p_2\inn V\inn p_3\right)\nonumber\\
&&+2p_2\inn V\inn p_2(p_2\inn V\inn p_3 \cJ_{1} -p_2\inn N\inn p_3 \cJ_{2})-2(-2 \cJ+\cJ_{13}+\cJ_{14}) p_2\inn V\inn p_3\,p_2\inn N\inn p_3=0\nonumber
\eeqa
where $\cI_7(p_1,p_2,p_3)=\cI_4(p_1,p_3,p_2)$. The relation in the first line has been appeared in the amplitude considered in \cite{Garousi:2010bm}. From the $(2\leftrightarrow 3)$ part of the  amplitude \reef{totalpole}, one finds the following relations:
\beqa
&&-2p_1\inn N\inn p_3\cI_1+2p_3\inn V\inn p_3\cI_4+p_2\inn N\inn p_3\cI_2-p_2\inn V\inn p_3\cI_3=0\labell{iden2}\\
&&-2 \cI_3 p_1\inn N\inn p_3+(\cJ_{13}-\cJ_{14}) p_2\inn N\inn p_3+2 \cJ_{16} p_3\inn V\inn p_3\nonumber\\
&&\qquad\qquad\qquad\qquad+(-4 \cJ+\cJ_{13}+\cJ_{14}+2 \cJ_{5}) p_2\inn V\inn p_3=0\nonumber\\
&&2 \cI_{2} p_1\inn N\inn p_3-2 \cJ_{4} p_3\inn V\inn p_3+(\cJ_{13}-\cJ_{14}) p_2\inn V\inn p_3\nonumber\\
&&\qquad\qquad\qquad\qquad+(\cJ_{13}+\cJ_{14}-2 \cJ_{5}) p_2\inn N\inn p_3=0\nonumber\\
&&-2 \cI_{7} p_1\inn N\inn p_3+\cJ_{2} p_2\inn N\inn p_3+2 \cJ_{3} p_3\inn V\inn p_3-\cJ_{1} p_2\inn V\inn p_3=0\nonumber\\
&&(-\cJ_{13}+\cJ_{14}) \left((p_2\inn N\inn p_3)^2+(p_2\inn V\inn p_3)^2\right)+2 p_1\inn N\inn p_3 \left(\cI_{3} p_2\inn N\inn p_3-\cI_{2} p_2\inn V\inn p_3\right)\nonumber\\
&&+2p_3\inn V\inn p_3(p_2\inn V\inn p_3 \cJ_{4} -p_2\inn N\inn p_3 \cJ_{12})-2(-2 \cJ+\cJ_{13}+\cJ_{14}) p_2\inn V\inn p_3\,p_2\inn N\inn p_3=0\nonumber
\eeqa
The relation in the first line has been also appeared in the amplitude considered in \cite{Garousi:2010bm}. If the explicit form of the integrals were known, then one could verify the above relations explicitly. We will  verify the above relations  for a special case in  which the integrals can be calculated explicitly.

An indirect check of the above relations is that using them one can write the amplitude in terms of  RR  field strength which can then be checked with the S-matrix element in $(-1/2,-1/2)$-picture in which the RR vertex operator is in terms of field strength $F$. Using the above relations, one can write \reef{totalpole} as 
\beqa
{\cal A}&\sim&\frac{1}{2}p_2^{a_0}p_3^{a_1}\bigg[
\bigg(
p_2\inn N\inn p_3(\veps_2\inn V\inn\veps_3)^{a_2a_3}-p_2\inn V\inn p_3(\veps_2\inn N\inn\veps_3)^{a_2a_3}-\veps_3^{a_2a_3}p_3\inn V\inn\veps_2\inn N\inn p_3\bigg)\cJ\nonumber\\
&&+\veps_3^{a_2a_3}\bigg(p_3\inn V\inn\veps_2\inn V\inn p_2\cJ_1-\frac{1}{2}p_3\inn V\inn\veps_2\inn N\inn p_1\cI_3
+p_2\inn V\inn\veps_2\inn N\inn p_3\cJ_2-p_3\inn V\inn\veps_2\inn N\inn p_3\cJ_5\nonumber\\
&&+\frac{1}{2}p_3\inn N\inn\veps_2\inn N\inn p_1\cI_2\bigg)-p_1\inn N\inn\veps_2^{a_2}\,p_2\inn V\inn\veps_3^{a_3}\cI_3-p_1\inn N\inn\veps_2^{a_2}\,p_1\inn N\inn\veps_3^{a_3}\cI_1\nonumber\\
&&+p_1\inn N\inn\veps_2^{a_2}\,p_2\inn N\inn\veps_3^{a_3}\cI_2+2p_3\inn V\inn\veps_2^{a_2}\,p_2\inn N\inn\veps_3^{a_3}\cJ_5+4p_1\inn N\inn\veps_2^{a_2}\,p_3\inn V\inn\veps_3^{a_3}\cI_4\nonumber\\
&&-2p_2\inn V\inn\veps_2^{a_2}\,p_2\inn N\inn\veps_3^{a_3}\cJ_2+2p_2\inn V\inn\veps_2^{a_2}\,p_2\inn V\inn\veps_3^{a_3}\cJ_1-4p_2\inn V\inn\veps_2^{a_2}\,p_3\inn V\inn\veps_3^{a_3}\cJ_3\bigg]\eps_{a_0a_1a_2a_3}\nonumber\\
&&+\frac{1}{2}p_1^{a_0}\bigg[\frac{1}{4}p_3\inn V\inn\veps_2^{a_2}\veps_3^{a_1a_3}\left(2p_2\inn V\inn p_2\cJ_1+2p_3\inn V\inn p_3\cJ_4-4p_2\inn N\inn p_3\cJ\right)\nonumber\\
&&+\frac{1}{4}p_3\inn N\inn\veps_2^{a_2}\veps_3^{a_1a_3}\left(\frac{}{}2(\cJ_{13}-\cJ_{14})p_2\inn N\inn p_3+(-4\cJ+\cJ_{13}+\cJ_{14})p_2\inn V\inn p_3\right)\nonumber\\
&&-2p_2\inn V\inn\veps_2^{a_2}\veps_3^{a_1a_3}p_3\inn V\inn p_3\cJ_3+p_1\inn N\inn\veps_2^{a_2}\veps_3^{a_1a_3}p_3\inn V\inn p_3\cI_4\bigg]\eps_{a_0a_1a_2a_3}\nonumber\\
&&+\frac{1}{2}(p_1)_i\bigg[\frac{1}{2}p_3\inn V\inn\veps_2^{a_2}\veps_3^{a_1a_3}(p_2^ip_2^{a_0}\cI_3+p_3^ip_3^{a_0}\cI_2)+\frac{1}{2}p_3\inn N\inn\veps_2^{a_2}\veps_3^{a_1a_3}(p_3^ip_2^{a_0}\cI_3+p_2^ip_3^{a_0}\cI_2)\nonumber\\
&&-2p_3^ip_3^{a_0}p_2\inn V\inn\veps_2^{a_2}\veps_3^{a_1a_3}\cI_7+p_3^ip_3^{a_0}p_1\inn N\inn\veps_2^{a_2}\veps_3^{a_1a_3}\cI_1\bigg]\eps_{a_0a_1a_2a_3}+(2\leftrightarrow 3)\nonumber\\
&&+2\veps_2^{a_0a_1}\veps_3^{a_2a_3}(p_1)_i\bigg[p_2^ip_3\inn V\inn p_3\cI_4+\frac{1}{2}p_3^ip_2\inn V\inn p_3\cI_2-\frac{1}{2}p_3^ip_2\inn N\inn p_3\cI_3\bigg]\eps_{a_0a_1a_2a_3}\labell{F}
\eeqa 
As we mentioned before, the result for the scattering amplitude can easily be extended to the arbitrary RR potential by replacing $\eps_{a_0a_1a_2a_3}$ with $\eps_{a_0\cdots a_p}\veps_1^{a_4\cdots a_p}/(p-4)!$ where $\veps_1^{a_4\cdots a_p}$ is the RR polarization tensor. The couplings in the last three lines above are consistent with the couplings found in \cite{Garousi:2010bm} for the RR potential with one transverse index. They can be combined to be written in terms of RR field strength $F_{ia_4\dots a_p}$. This part of amplitude has been checked explicitly in \cite{Garousi:2010bm} by the evaluation of the S-matrix element in $(-1/2,-1/2)$-picture.  The other couplings can easily be written in terms of $F_{a_0a_4\cdots a_p}$. We confirmed them by evaluating the S-matrix element in $(-1/2,-1/2)$-picture.

Having written the amplitude in terms of the RR field strength, one observes that 
the amplitude at each order of  $\alpha'$  enjoys the RR gauge symmetry, as expected. In particular, this symmetry appear in the amplitude at order $O(\alpha'^2)$ which, as we will see,  includes contact terms, massless open and closed string poles.

We now try to write the amplitude \reef{totalpole} in terms of $H$. As we mentioned before, strategy for doing this step if to look at the amplitude \reef{total} and find terms which are not proportional to the Mandelstam variables and write them in terms of $H$ and some extra terms which are proportional to the Mandelstam variables. Then using the  relations \reef{iden3} and \reef{iden2}, one should simplify the terms which are proportional to the Mandelstam variables. The remaining terms which are proportional to the Mandelstam variables, should then be either zero or be written in terms of $H$. Doing this, one finds the following result:
\beqa
{\cal A}&\!\!\!\!\!\sim\!\!\!\!\!&\bigg[(p_2)_aH_2^{aa_0a_1}(p_3)_bH_3^{ba_2a_3}\cJ_3-\frac{1}{2}(p_2)_a(p_2)_bH_2^{aa_0a_1}H_3^{ba_2a_3}\cJ_1+\frac{1}{2}(p_2)_a(p_2)_iH_2^{aa_0a_1}H_3^{ia_2a_3}\cJ_2\nonumber\\
&\!\!\!\!\!\!\!\!\!\!\!\!\!\!\!\!\!\!\!\!\!\!\!\!&-(p_1)_i(p_2)_aH_2^{aa_0a_1}H_3^{ia_2a_3}\cI_7-\frac{1}{2}(p_2)_iH_2^{aa_0a_1}(p_3)_aH_3^{ia_2a_3}\cJ_5-\frac{1}{4}(p_1)_i(p_2)_jH_2^{ia_0a_1}H_3^{ja_2a_3}\cI_2\nonumber\\
&\!\!\!\!\!\!\!\!\!\!\!\!\!\!\!\!\!\!\!\!\!\!\!\!&+\frac{1}{4}(p_1)_i(p_1)_jH_2^{ia_0a_1}H_3^{ja_2a_3}\cI_1+\frac{1}{4}(p_1)_i(p_2)_aH_2^{ia_0a_1}H_3^{aa_2a_3}\cI_3-\frac{1}{3}(p_2)_aH_2^{a_0a_1a_2}(p_3)_bH_3^{aba_3}\cJ_4\nonumber\\
&\!\!\!\!\!\!\!\!\!\!\!\!\!\!\!\!\!\!\!\!\!\!\!\!&-\frac{1}{6}(p_1)_i(p_2)_aH_2^{a_0a_1a_2}H_3^{iaa_3}\cI_2+\frac{1}{3}(p_2)_iH_2^{a_0a_1a_2}(p_3)_aH_3^{iaa_3}\cJ_{12}+\frac{1}{6}(p_1)_i(p_2)_jH_2^{a_0a_1a_2}H_3^{ija_3}\cI_3\nonumber\\
&\!\!\!\!\!\!\!\!\!\!\!\!\!\!\!\!\!\!\!\!\!\!\!\!&-\frac{1}{3}(p_2)_i(p_2)_aH_2^{a_0a_1a_2}H_3^{iaa_3}(-\cJ_{5}+\cJ)\nonumber\\
&\!\!\!\!\!\!\!\!\!\!\!\!\!\!\!\!\!\!\!\!\!\!\!\!&+\frac{1}{4}(p_2\inn N\inn p_3)H_2^{aa_0a_1}H_3^{aa_2a_3}\cJ-\frac{1}{4}(p_2\inn V\inn p_3)H_2^{ia_0a_1}H_3^{ia_2a_3}\cJ\bigg]\frac{1}{2}\eps_{a_0a_1a_2a_3}+\bigg[2\leftrightarrow 3\bigg]\labell{2H}
\eeqa
The terms in the last line are the only terms which are proportional to the Mandelstam variables. This is our final result for the string theory scattering amplitude. We will fix the normalization of the amplitude in section 2.2 by comparing the above amplitude at  low energy   with the corresponding field theory. The contracted indices in  the terms in the last line are not momentum indices, so they are not invariant under the linear T-duality. They combine with the corresponding  terms in the gravity amplitude \reef{gg}   to produce a T-dual amplitude.  All other terms are invariant under the linear T-duality. Hence, the combination of \reef{2H} and \reef{gg} is invariant under the linear T-duality when the Killing coordinate is an index of the RR potential. In other cases, one should add the amplitude for higher RR potential in which we are not interested in this paper.

Note that the above amplitude is not in terms of RR field strength. 
Hence, the amplitude can be written either in terms of H or in terms of the RR field strength. This indicates that the field theory couplings which are invariant under the B-field gauge transformations, are not invariant under the RR gauge transformation. However, as we mentioned before, the combination of the field theory couplings and the massless open and closed string poles at each order of $\alpha'$ is invariant under the RR gauge transformation.

\subsubsection{Low energy limit}

To find the low energy limit of the string theory amplitude \reef{2H}, we are now trying  to evaluate the integrals. It is hard to evaluate the integrals for the general case, so we concentrate on the  special kinematic setup \cite{Garousi:2010er,Garousi:2010bm}.  Examining  the Feynman diagrams involved, one can easily verify that   the amplitude considered in this paper has no massless pole in the $p_2\inn p_3$-channel. 
Moreover,  there is no closed or open string channel corresponding to the Mandelstam variable $p_2\inn D\inn p_3$, hence,   
 we restrict the Mandelstam variables  to 
\beqa
 p_2\inn D\inn p_3=0, &{\rm and }& 
 p_2\inn p_3=0\labell{kin}
 \eeqa
Even though the amplitude has no massless pole in $p_2\inn p_3$-channel, the integrals which appear with the coefficient $p_2\inn p_3$ in the amplitude may have massless pole in $p_2\inn p_3$, so one can not set to zero the terms which are proportional to $p_2\inn p_3$. The integrals $\cI_2,\, \cI_4$, $ \cJ_1,\,\cJ_2$, $\cJ_3,\, \cJ_5$ and $\cJ_{14}$ which appear in the amplitude \reef{totalpole} have no  pole in $p_2\inn p_3$. The reason for this is that if one uses the constraint \reef{kin} the result of integrals would  be finite. However, the the result of integral $\cJ_{13}$ under the constraint \reef{kin} is infinite, hence, it has massless pole in $p_2\inn p_3$.  

Using the Maple, one can easily  preform the $\theta $-integral in the integrals $\cI_1,\,\cI_2$, $ \cI_4,\, \cJ_1$, $\cJ_2,\,\cJ_3$, $ \cJ_5$ which appear in the amplitude \reef{2H}, for the constraint \reef{kin}. The result is
\beqa
\cI_1&=&-2\pi\int_0^1dr_3\int_0^{1}dr_2\frac{\tK'}{r_2r_3}\nonumber\\
\cI_2&=&4\pi\int_0^1dr_3\int_0^{r_3}dr_2\frac{\tK'}{r_2r_3}\nonumber\\
\cI_4&=&-2\pi\int_0^1dr_2\int_0^{1}dr_3\frac{(1+r_3^2)\tK'}{r_2r_3(1-r_3^2)}\labell{integ}\\
\cJ_1&=&4\pi\int_0^1dr_2\int_0^{r_2}dr_3\frac{(1+r_2^2)\tK'}{r_2r_3(1-r_2^2)}\nonumber\\
\cJ_2&=&4\pi\int_0^1dr_3\int_0^{r_3}dr_2\frac{(1+r_2^2)\tK'}{r_2r_3(1-r_2^2)}\nonumber\\
\cJ_3&=&-2\pi\int_0^1dr_2\int_0^{1}dr_3\frac{(1+r_3^2)(1+r_2^2)\tK'}{r_2r_3(1-r_3^2)(1-r_2^2)}\nonumber\\
\cJ_5&=&2\pi\int_0^1dr_2\int_0^{r_2}dr_3\frac{(1+r_3^2)\tK'}{r_2r_3(1-r_3^2)}-(2\leftrightarrow 3)\nonumber
\eeqa
where $\tK'$ is the value of $\tK$ in the constraint \reef{kin}, \ie
\beqa
\tK'&=&{r_2}^{2p_1\cdot p_2}\ {r_3}^{2 p_1\cdot p_3} (1-{r_2}^2)^{p_2\cdot D\cdot p_2}(1-{r_3}^2)^{p_3\cdot D\cdot p_3}\labell{K'}
\eeqa
Using the definition of beta function 
\beqa
\int_0^1dx\,x^{\alpha-1}(1-x)^{\beta-1}&=&B(\alpha,\beta)
\eeqa
The radial integral in $\cI_1,\, \cI_4$ and $\cJ_3$ becomes
\beqa
\cI_1&=&-\frac{\pi}{2} B(s, 1+p)B(t, 1+q)\nonumber\\
\cI_4&=&-\frac{\pi}{2} \frac{(2t+q)}{q}B(s, 1+p)B(t, 1+q)\labell{integ147}\\
\cJ_{3}&=&-\frac{\pi}{2} \frac{(2s+p)}{p}\frac{(2t+q)}{q}B(s, 1+p)B(t, 1+q)\nonumber
\eeqa
where we have used the following definitions for the Mandelstam variables:
\beqa
s=p_1\inn p_2&;&t=p_1\inn p_3\nonumber\\
p=p_2\inn D\inn p_2&;&q=p_3\inn D\inn p_3
\eeqa

The radial integrals in $\cI_2,\,\cJ_1,\,\cJ_2$ and $\cJ_5$ have the following structure:
\beqa
I&=&\int_0^1dx\int_0^xdy\,x^ay^b(1-x)^c(1-y)^d
\eeqa
which has the solution (see the appendix in \cite{Garousi:2010bm})
\beqa
I&=&\frac{B(1+c,2+a+b)}{1+b}{}_3F_2\bigg[{2+a+b,\ 1+b, \ -d \atop 3+a+b+c, \ 2+b}\ ;\ 1\bigg]\labell{II}
\eeqa
Using this formula, one finds
\beqa
\cI_2&=&\pi\frac{B(s+t,1+q)}{s}{}_3F_2\bigg[{s,\ s+t, \ -p \atop 1+s, \ 1+s+t+q}\ ;\ 1\bigg]\nonumber\\
\cJ_1&=&\frac{\pi}{t}\left(B(s+t, p){}_3F_2\bigg[{t,\ s+t, \ -q \atop 1+t, \ s+t+p}\ ;\ 1\bigg]\right.\nonumber\\
&&\left.+B(1+s+t, p){}_3F_2\bigg[{t,\ 1+s+t, \ -q \atop 1+t, \ 1+s+t+p}\ ;\ 1\bigg]\right)\nonumber\\
\cJ_2&=&\pi\left(\frac{B(s+t, 1+q)}{s}{}_3F_2\bigg[{s,\ s+t, \ 1-p \atop 1+s, \ 1+s+t+q}\ ;\ 1\bigg]\right.\nonumber\\
&&\left.+\frac{B(1+s+t,1+q)}{1+s}{}_3F_2\bigg[{1+s,\ 1+s+t, \ 1-p \atop 2+s, \ 2+s+t+q}\ ;\ 1\bigg]\right)\nonumber\\
\cJ_5&=&\pi\left(\frac{B(s+t, 1+p)}{t}{}_3F_2\bigg[{t,\ s+t, \ 1-q \atop 1+t, \ 1+s+t+p}\ ;\ 1\bigg]\right.\nonumber\\
&&\left.+\frac{B(1+s+t,1+p)}{1+t}{}_3F_2\bigg[{1+t,\ 1+s+t, \ 1-q \atop 2+t, \ 2+s+t+p}\ ;\ 1\bigg]\right)-(2\leftrightarrow 3)\nonumber
\eeqa
The evaluation of the integrals $\cJ_{13}$ and $\cJ_{14}$ is presented in the appendix A.

Having found the explicit form of the integrals for the constraint kinematic setup \reef{kin}, we now verify the relations between the integrals in \reef{iden3}.  Since none of the integrals have simple pole at $p_2\inn p_3$ or $p_2\inn D\inn p_3$, except $\cJ_{13}$ which has only simple massless pole at $p_2\inn p_3$ (see appendix A), the relations \reef{iden3} simplify to 
\beqa
&&-2p_1\inn N\inn p_2\cI_1+2p_2\inn V\inn p_2\cI_7=0\labell{iden33}\\
&&-2 \cI_2 p_1\inn N\inn p_2+\cJ_{13} p_2\inn p_3+2 \cJ_2 p_2\inn V\inn p_2=0\nonumber\\
&&2 \cI_{3} p_1\inn N\inn p_2-2 \cJ_{1} p_2\inn V\inn p_2+\cJ_{13} p_2\inn  p_3=0\nonumber\\
&&-2 \cI_{4} p_1\inn N\inn p_2+2 \cJ_{3} p_2\inn V\inn p_2=0\nonumber
\eeqa
Note that $\cJ_{13} (p_2\inn  p_3)^2$ is zero whereas $\cJ_{13} p_2\inn  p_3$ is nonzero. Using the equation \reef{integ147}, one can easily verify the first and last relations. The subtraction of the second and the third relations give
\beqa
&&-2p_1\inn N\inn p_2(\cI_2+\cI_3)+2p_2\inn V\inn p_2(\cJ_1+\cJ_2)=0\nonumber
\eeqa
Using the integral representations in \reef{integ}, one observes that $\cI_2+\cI_3=-2\cI_1$ and $\cJ_1+\cJ_2=-2\cI_7$. Hence the above relation reduces to the first relation in \reef{iden33}.
 
 To study the low energy limit of the amplitude \reef{2H}, we expand $\cI_1,\,\cI_2,\cI_4, \cJ_1,\cJ_2,\cJ_3$ and $\cJ_{5}$  at the low energy. The expansion of beta function is standard and for expanding the hypergeometric function we use the package \cite{Huber:2005yg}. The result is  
\beqa
\cI_1&=&-\frac{\pi}{2}\left(\frac{1}{s(s+t)}+\frac{1}{t(s+t)}-\frac{\pi^2}{6}\left(\frac{q}{s}+\frac{p}{t}\right)+\cdots\right)\nonumber\\
\cI_2&=&\pi\left(\frac{1}{s(s+t)}-\frac{\pi^2}{6}\frac{q}{s}+\cdots\right)\labell{expand}\\
\cI_4&=&-\frac{\pi}{2}\left(\frac{(2t+q)}{q}\right) \left(\frac{1}{st}-\frac{\pi^2}{6}\bigg[\frac{p}{t}+\frac{q}{s}\bigg]+\cdots\right)\nonumber\\
\cJ_1&=&\pi\left(\frac{1}{t(s+t)}+\frac{2}{pt}-\frac{\pi^2}{6}\bigg[2+\frac{(2s+p)}{t}+\frac{2q}{p}\bigg]+\cdots\right)\nonumber\\
\cJ_2&=&\pi\left(\frac{1}{s(s+t)}+\frac{\pi^2}{6}\bigg[2-\frac{q}{s}\bigg]+\cdots\right)\nonumber\\
\cJ_3&=&-\frac{\pi}{2}\left(\frac{(2s+p)}{p}\frac{(2t+q)}{q}\right) \left(\frac{1}{st}-\frac{\pi^2}{6}\bigg[\frac{p}{t}+\frac{q}{s}\bigg]+\cdots\right)\nonumber\\
\cJ_5&=&\pi\left(\frac{1}{t(s+t)}-\frac{1}{s(s+t)}+\frac{\pi^2}{6}\bigg[\frac{q}{s}-\frac{p}{t}\bigg]+\cdots\right)\nonumber
\eeqa

 From these expansions and the expansion \reef{exp12} for $\cJ$, one finds the following contact terms at order $O(\alpha'^2)$:
 \beqa
{\cal A}^{contact}&\sim&\frac{\pi^3}{6}\bigg[\frac{1}{2}(p_2)_a(p_2)_bH_2^{aa_0a_1}H_3^{ba_2a_3}+\frac{1}{2}(p_2)_a(p_2)_iH_2^{aa_0a_1}H_3^{ia_2a_3}\nonumber\\
&&+\frac{1}{3}(p_2)_aH_2^{a_0a_1a_2}(p_3)_bH_3^{aba_3}+\frac{1}{3}(p_2)_iH_2^{a_0a_1a_2}(p_3)_aH_3^{iaa_3}\nonumber\\
&&-\frac{1}{4}(p_2)_iH_2^{aa_0a_1}(p_3)_iH_3^{aa_2a_3}+\frac{1}{4}(p_2)_aH_2^{ia_0a_1}(p_3)_aH_3^{ia_2a_3}\nonumber\\
&&\qquad\qquad\qquad+\frac{1}{3}(p_2)_i(p_2)_aH_2^{a_0a_1a_2}H_3^{iaa_3}\bigg]\eps_{a_0a_1a_2a_3}+\bigg[2\leftrightarrow 3\bigg]\labell{contact}
\eeqa
 Even though we have found the expansion \reef{expand} for the constraint \reef{kin}, the constants of the integrals which produce the above contact terms, are independent of $p_2\inn p_3$ or $\p_2\inn D\inn p_3$. Hence, the above result is valid for the general case. 

The amplitude has also  the following massless open string poles at order $O(\alpha'^2)$:
\beqa
{\cal A}^{ pole}&\sim&\frac{\pi^3}{12}\bigg[2(p_2)_aH_2^{aa_0a_1}(p_3)_bH_3^{ba_2a_3}\left(\frac{q+2t}{p}\right)-(p_1)_i(p_2)_aH_2^{aa_0a_1}H_3^{ia_2a_3}\frac{q}{p}\labell{pole}\\
&&
+\frac{2}{3}(p_2)_bH_2^{aba_3}(p_3)_aH_3^{a_0a_1a_2}\frac{q}{p}+(p_2)_a(p_2)_bH_2^{aa_0a_1}H_3^{ba_2a_3}\frac{q}{p}\bigg]\eps_{a_0a_1a_2a_3}+\bigg[2\leftrightarrow 3\bigg]\nonumber
\eeqa
  Since the expansion \reef{expand} for the integrals are valid for the constraint \reef{kin}, there might be some other massless open string poles which are proportional to $p_2\inn V\inn p_3$ or $p_2\inn N\inn p_3$. 

The amplitude \reef{2H} has also some massless closed string poles  at order $O(\alpha'^2)$ which are in terms of $H$. However, they are not invariant under the RR gauge transformations. Recall that the  RR gauge transformation of $(n-2)$-form potential in supergravity  is  canceled with the gauge transformation of $n$-form. So one does not expect to have Ward identity corresponding to  a RR potential in the massless closed string poles. We have seen that the string amplitude \reef{2H} can be written in terms of RR field strength, \ie  \reef{F}. So this amplitude at order $O(\alpha'^2)$ which is  equal to the sum of the above massless closed string amplitude, massless open string amplitude \reef{pole} and the contact terms \reef{contact}, is invariant under the RR gauge transformation.  We are not interested in the massless closed string poles, as they do not produce  any new D-brane couplings.  
 
\begin{center} \begin{picture}(100,100)(0,0)
\SetColor{Black}
\Line(10,40)(10,90)
\Line(10,90)(30,110)
\Line(30,110)(30,60)
\Line(30,60)(10,40)
\Text(0,108)[]{$D_p$-brane}
\Vertex(20,90){1.5}\Vertex(20,60){1.5}
\Photon(20,60)(20,90){2}{7} \Text(12,75)[]{$A$}
\Gluon(20,90)(65,90){3}{4} \Text(42.5,102.5)[]{$B^{(2)}$}
\Gluon(20,60)(65,60){3}{4}\Text(42.5,72.5)[]{$B^{(2)}$}
\Photon(20,60)(50,35){3}{4} \Text(63,45)[]{$C^{(p-3)}$}
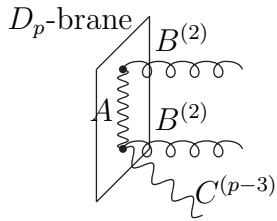
\end{picture} \\ {\sl Figure 1: {\rm  Feynman diagram for massless open string poles.}}
\end{center}
\subsection{Field theory amplitude}

Having found the contact terms and the massless open string poles of string amplitude at order $O(\alpha'^2)$, we now reproduce them by appropriate couplings in field theory. The field theory couplings are those that we have already presented in section 1. To simplify the calculation we consider the  RR scalar. In the next subsection we show that the field theory produces the  the massless open string poles \reef{pole}. 

\subsubsection{Open string pole}

In  field theory, the open string channel of the scattering amplitude of one RR potential $(p-3)$-form and two B-fields  is given by the Feynman diagram in fig.1. The corresponding Feynman  amplitude is given by:
\beqa
{\cal A}_1^f&=&V_a(\veps_3,A)G_{ab}(A) V_b(A,\veps_2,\veps_1^{(p-3)})+(2\leftrightarrow 3)\labell{Af}
\eeqa
where $A^a$ is the gauge field on the D$_p$-brane. 
The gauge field propagator and the vertex $V_a(\veps_3,A)$ can be read from the DBI action \reef{DBI}, \ie
\beqa
V_a(\veps_3,A)&=&(2\pi\alpha')T_3(p_3\inn V\inn\veps_3)_a\nonumber\\
G_{ab}(A)&=&\left(\frac{-i}{T_3(2\pi\alpha')^2}\right)\frac{\eta_{ab}}{p_3\inn V\inn p_3}
\eeqa
 
The   couplings in section 1 produce the vertex $ V_b(A,\veps_2,\veps_1^{(p-3)})$.  We begin by considering the couplings in the second line of \reef{77} for RR scalar. The  vertex corresponding to the first term is given by  
\beqa
V_{b}(A,\veps_2)&=&-2(\pi\alpha')^3T_3\eps_{a_0a_1 a_2b}(p_1+p_2)^{a_2}(p_2)_{a}\bigg[H_2^{a_0a_1a}p_1\inn N\inn p_2\bigg]\nonumber
\eeqa
The amplitude \reef{Af} then becomes
\beqa
{\cal A}_1^f&=&i(\pi\alpha')^2T_3\frac{(p_3\inn V\inn\veps_3)^{a_3}p_3^{a_2}}{p_3\inn V\inn p_3}\eps_{a_0\cdots a_3}(p_2)_{a}\bigg[H_2^{a_0a_1a}p_1\inn N\inn p_2\bigg]+(2\leftrightarrow 3)\nonumber
\eeqa
This amplitude is of order $O(\alpha'^2)$ which has  six momentum in the numerator and two momentum in the denominator. 
The couplings in the second line of \reef{77} have also  the following  contact term: 
\beqa
{\cal A}_2^f&=&i\left(\frac{(\pi\alpha')^2T_3}{2}\right)\veps_3^{a_2a_3}\eps_{a_0\cdots a_3}(p_2)_{a}\bigg[H_2^{a_0a_1a}p_1\inn N\inn p_2\bigg]+(2\leftrightarrow 3)\labell{af1}
\eeqa
Using the following identity:
\beqa
\eps_{a_0\cdots a_3}(p_3\inn V\inn H_3)^{a_2a_3}&=&\eps_{a_0\cdots a_3}\left(2p_3\inn V\inn\veps_3^{a_2}p_3^{a_3}+p_3\inn V\inn p_3\veps_3^{a_2a_3}\right)\nonumber
\eeqa
one can rewrite the sum of ${\cal A}_1^f$ and ${\cal A}_2^f$ as
\beqa
{\cal A}^f&=&i\left(\frac{(\pi\alpha')^2T_3}{2}\right)\frac{1}{p_3\inn V\inn p_3}\eps_{a_0\cdots a_3}(p_2)_{a}(p_3)_b\bigg[H_2^{a_0a_1a} H_3^{ba_2a_3}p_1\inn N\inn p_2\bigg]+(2\leftrightarrow 3)\labell{af2}
\eeqa
Using the constraint \reef{kin}, one writes $p_1\inn N\inn p_2=s+p/2$. Hence, the above result  is exactly the first term in the open string  amplitude \reef{pole} provided that one fix the normalization of the string amplitude \reef{2H} to be
\beqa
N&=&\frac{3i\alpha'^2T_p}{\pi}\labell{N}
\eeqa 

Now consider the second term  in the second line of \reef{77}. It produces the following  vertex and contact term:
\beqa
V_{b}(A,\veps_2)&=&(\pi\alpha')^3T_3\eps_{a_0a_1 a_2b}(p_1+p_2)^{a_2}(p_1)_{i}\bigg[H_2^{a_0a_1i}p_2\inn V\inn p_2\bigg]\nonumber\\
{\cal A}_2^f&=&-i\left(\frac{(\pi\alpha')^2T_3}{4}\right)\veps_3^{a_2a_3}\eps_{a_0\cdots a_3}(p_1)_{i}\bigg[H_2^{a_0a_1i}p_2\inn V\inn p_2\bigg]+(2\leftrightarrow 3)\nonumber
\eeqa
Doing the same steps as before, one finds
\beqa
{\cal A}^f&=&-i\left(\frac{(\pi\alpha')^2T_3}{4}\right)\frac{1}{p_3\inn V\inn p_3}\eps_{a_0\cdots a_3}(p_1)_{i}(p_3)_a\bigg[H_2^{a_0a_1i} H_3^{aa_2a_3}p_2\inn V\inn p_2\bigg]+(2\leftrightarrow 3)\labell{Af2}
\eeqa
This  is  exactly the second term in the open string  amplitude \reef{pole}.

Next consider the first two terms in the last line of \reef{5}. They produce the following vertex and contact term:
\beqa
V_{b}(A,\veps_2)&=&-\frac{(\pi\alpha')^3T_3}{3}p_2\inn V\inn p_2H_2^{a_0a_1a_2}\bigg[\eps_{a_0a_1 a_2b}(p_1+p_2)\inn V\inn p_2-\eps_{a_0a_1 a_2a_3}(p_1+p_2)^{a_3}(p_2)_b\bigg]\nonumber\\
{\cal A}_2^f&=&i\left(\frac{(\pi\alpha')^2T_3}{6}\right)p_2\inn V\inn p_2H_2^{a_0a_1a_2}\eps_{a_0\cdots a_3}p_2\inn V\inn\veps_3^{a_3}+(2\leftrightarrow 3)\nonumber
\eeqa
In this case, one finds
\beqa
{\cal A}^f&=&-i\left(\frac{(\pi\alpha')^2T_3}{6}\right)\frac{1}{p_3\inn V\inn p_3}\eps_{a_0\cdots a_3}(p_3)_{a}(p_2)_b\bigg[H_2^{a_0a_1a_2} H_3^{aba_3}p_2\inn V\inn p_2\bigg]+(2\leftrightarrow 3)\labell{Af3}
\eeqa
which is exactly the first term in the second line of \reef{pole}.

Finally consider the last two terms in the last line of \reef{5}. They produce the following vertex and contact term:
\beqa
V_{b}(A,\veps_2)&=&-(\pi\alpha')^3T_3p_2\inn V\inn p_2H_2^{a_0a_1a}(p_1+p_2)_a(p_1+p_2)^{a_2}\eps_{a_0a_1 a_2b}\nonumber\\
{\cal A}_2^f&=&i\left(\frac{(\pi\alpha')^2T_3}{4}\right)p_2\inn V\inn p_2H_2^{a_0a_1a}\veps_3^{a_2a_3}(p_3)_a\eps_{a_0\cdots a_3}+(2\leftrightarrow 3)\nonumber
\eeqa
In this case, one finds
\beqa
{\cal A}^f&=&i\left(\frac{(\pi\alpha')^2T_3}{4}\right)\frac{1}{p_3\inn V\inn p_3}\eps_{a_0\cdots a_3}(p_3)_{a}(p_3)_b\bigg[H_2^{a_0a_1a} H_3^{ba_2a_3}p_2\inn V\inn p_2\bigg]+(2\leftrightarrow 3)\labell{Af4}
\eeqa
which is exactly the last term in the second line of \reef{pole}.

The above calculation indicates that the couplings which include  $(B+2\pi\alpha' f)$ appears as massless open string pole. Since the massless open string ampliude \reef{pole} does not includes terms which are proportional to $p_2\inn V\inn p_3$ or $p_2\inn N\inn p_3$, the above calculation can not fix the coefficient of all    higher derivative couplings which contain  $(B+2\pi\alpha' f)$. For example  the higher derivative  coupling $C\wedge\prt_a\prt_b f\wedge\prt^a\prt^b f$  can be read from the S-matrix element of one RR and two gauge field vertex operators \cite{Hashimoto:1996kf,Garousi:1998fg}. One may extend it to $C\wedge\prt_a\prt_b (B+2\pi\alpha' f)\wedge\prt^a\prt^b (B+2\pi\alpha' f)$. This coupling produces massless open string pole which is proportional to $p_2\inn V\inn p_3$. Hence, our calculation can not fix the presence of such couplings. 

The couplings corresponding to the massless open string poles can also be extracted from the low energy limit of the S-matrix element of one RR, one B-field and one open string gauge field vertex operators. In that case, after fixing the $SL(2,R)$ symmetry, one  would find a double integral which can be evaluated explicitly for the general kinematic setup \cite{Fotopoulos:2001pt}. 

\subsubsection{Contact terms}

Using the normalization \reef{N}, one finds the Lagrangian corresponding to the contact terms in \reef{contact} to be
\beqa
{\cal L}&=&\frac{(\pi\alpha')^2T_3}{2}C^{(0)}\bigg[\frac{1}{2}H^{aa_0a_1,ab}H^{ba_2a_3}+\frac{1}{2}H^{aa_0a_1,ai}H^{ia_2a_3}+\frac{1}{3}H^{a_0a_1a_2,ia}H^{iaa_3}\nonumber\\
&&\qquad\qquad+\frac{1}{3}H^{a_0a_1a_2,a}H^{aba_3,b}+\frac{1}{3}H^{a_0a_1a_2,i}H^{iaa_3,a}\nonumber\\
&&\qquad\qquad-\frac{1}{4}H^{aa_0a_1,i}H^{aa_2a_3,i}+\frac{1}{4}H^{ia_0a_1,a}H^{ia_2a_3,a}\bigg]\eps_{a_0a_1a_2a_3}\labell{final}
\eeqa
The terms in the last line are exactly the couplings \reef{Tf41new}. All other terms are those appear in the first two lines of \reef{5}. Unlike the open string poles, there are  no other couplings at order $O(\alpha^2)$.

The couplings in the last line above have been found in \cite{Garousi:2009dj} by requiring the the Chern-Simons couplings \reef{Tf2} to be invariant under linear T-duality, and by requiring the new couplings to be  invariant  under B-field gauge transformation.  These  couplings,   however, are not invariant under linear T-duality if one of the indices of B-field which is contracted with the volume form is the Killing coordinate. In that case one should add new couplings involving higher RR potential to make a complete T-dual multiplet \cite{Garousi:2009dj}. The new couplings, however, are neither  covariant nor invariant under the B-field gauge transformation. So one needs to add some other T-dual multiplets \cite{Garousi:2009dj}. Having found the new couplings in the first two lines above, one should do the same steps for these couplings as well, to find all nonzero couplings of D$_p$-branes at order $O(\alpha'^2)$. 

Extending the  calculation of the S-matrix element in this paper to the case that $n=p-1$, $n=p+1$, $n=p+3$ and $n=p+5$, one would be able to find all nonzero couplings.   We have performed the calculation for $n=p+5$ case and found  that the S-matrix element is zero for two gravitons, and  has only closed string poles   for two B-fields. We present this result in the appendix B.

 {\bf Acknowledgments}:   This work is supported by Ferdowsi University of Mashhad under grant 2/16340-1389/10/14.   

\vspace*{1cm}

\newpage
{\bf\Large {A. Evaluating $\cJ_{13}$ and $\cJ_{14}$}}

In this appendix we calculate the integrals $\cJ_{13}$ and $\cJ_{14}$. These integrals do not appear in the amplitude \reef{2H}, however, they are needed to verifies the relations between the integrals that have been found in \reef{iden3} and \reef{iden2}.

The integral $\cJ_{14}$ has no pole in $p_2\inn p_3$. To take the $\theta$-integral in $\cJ_{14}$, we first write it as
\beqa
 \cJ_{14}&=&\cJ_{14}'+\cJ_{14}''\nonumber
 \eeqa
 where
 \beqa
 \cJ_{14}'&=&\int_0^1dr_2\int_0^{1}dr_3\frac{(r_2^2-r_3^2)^2}{r_2r_3(1-r_2^2)(1-r_3^2)}\int_0^{2\pi}d\theta\frac{\tK}{(r_2^2+r_3^2-2r_2r_3\cos(\theta))}\nonumber\\
\cJ_{14}''&=&\int_0^1dr_2\int_0^{1}dr_3\frac{(1+r_2^2r_3^2-2r_2^2)(1+r_2^2r_3^2-2r_3^2)}{r_2r_3(1-r_2^2)(1-r_3^2)}\int_0^{2\pi}d\theta\frac{\tK}{(1+r_2^2r_3^2-2r_2r_3\cos(\theta))}\nonumber
 \eeqa
 The $\theta$-integrals then become
\beqa
\cJ_{14}'&=&2\pi\int_0^1dr_2\int_0^{r_2}dr_3\frac{(r_2^2-r_3^2)\tK'}{r_2r_3(1-r_2^2)(1-r_3^2)}+(2\leftrightarrow 3)\nonumber\\
\cJ_{14}''&=&-\pi\int_0^1dr_2\int_0^{1}dr_3\frac{\tK'}{r_2r_3}\left(\frac{(r_2^2r_3^2-1)}{(1-r_2^2)(1-r_3^2)}+\frac{4}{(1-r_3^2)}+\frac{4}{(r_2^2r_3^2-1)}\right)+(2\leftrightarrow 3)\nonumber
\eeqa
Writing $r_2^2-r_3^2=(r_2^2-1)+(1-r_3^2)$ and using the formula \reef{II} the radial integral in $\cJ_{14}'$ becomes
\beqa
\cJ_{14}'&=&\frac{\pi}{2t}\left(B(s+t, p){}_3F_2\bigg[{t,\ s+t, \ -q \atop 1+t, \ s+t+p}\ ;\ 1\bigg]\right.\nonumber\\
&&\left.-B(s+t, 1+p){}_3F_2\bigg[{t,\ s+t, \ 1-q \atop 1+t, \ 1+s+t+p}\ ;\ 1\bigg]\right)+(2\leftrightarrow 3)\nonumber
\eeqa
The radial integral in the first two terms in $\cJ_{14}''$ gives a multiple of two beta functions. To take the radial integral in the last term we use the following identity:
\beqa
\frac{1}{1-r_2^2r_3^2}&=&{}_2F_1\bigg[{1,\ 1 \atop 1}\ ; \ r_2^2r_3^2\bigg]\labell{F21}
\eeqa
Then using the integral representation of the generalized hypergeometric function ${}_pF_q$ \cite{ISG}:
\beqa
\int_0^1dx\,x^a(1-x)^b{}_{p}F_{q}\bigg[{ a_1, \ \cdots, a_p\atop   b_1, \ \cdots, \ b_q}\ ;\ \la x
\bigg]=\nonumber\\
{}_{p+1}F_{q+1}\bigg[{1+a,\ a_1, \ \cdots, a_p\atop  2+a+b,\ b_1, \ \cdots, \ b_q}\ ;\ \la 
\bigg]B(1+a,1+b)\labell{Fpq}
\eeqa
one can write the radial integral  in terms of the the hypergeometric function ${}_{4}F_{3}$. The result is
\beqa
\cJ_{14}''&=&\frac{\pi}{4}\left(\frac{}{}B(s,p)B(t,q)-B(1+s,p)B(1+t,q)-4B(s,1+p)B(t,q)\right.\nonumber\\
&&\left.+4B(t,1+q)B(s,1+p){}_3F_2\bigg[{t,\ s, \ 1 \atop 1+t+q, \ 1+s+p}\ ;\ 1\bigg]\right)+(2\leftrightarrow 3)
\eeqa
where we have also used the identity:
\beqa
{}_4F_3\bigg[{a,\ b,\ c, \ 1 \atop d,\ e, \ 1}\ ;\ 1\bigg]&=&{}_3F_2\bigg[{a,\ b, \ c \atop d, \ e}\ ;\ 1\bigg]
\eeqa

The  integral $\cJ_{13}$, however,  has simple pole $1/(p_2\inn p_3)$.  To see this we write it as 
\beqa
\cJ_{13}&=&\cJ_{13}'+\cJ_{13}''\nonumber
\eeqa
where 
\beqa
\cJ_{13}'&=&-\int_0^1dr_2\int_0^{1}dr_3\frac{(1-r_2^2r_3^2)^2}{r_2r_3(1-r_2^2)(1-r_3^2)}\int_0^{2\pi}d\theta\frac{\tK}{(1+r_2^2r_3^2-2r_2r_3\cos(\theta))}\nonumber\\
\cJ_{13}''&=&\int_0^1dr_2\int_0^{1}dr_3\frac{(r_2^2+r_3^2-2)(2r_2^2r_3^2-r_2^2-r_3^2)}{r_2r_3(1-r_2^2)(1-r_3^2)}\int_0^{2\pi}d\theta\frac{\tK}{(r_2^2+r_3^2-2r_2r_3\cos(\theta))}\nonumber
\eeqa
 The $\theta$-integral in $\cJ_{13}'$ gives 
\beqa
\cJ_{13}'&=&-\pi\int_0^1dr_2\int_0^{1}dr_3\frac{(1-r_2^2r_3^2)\tK'}{r_2r_3(1-r_2^2)(1-r_3^2)}+(2\leftrightarrow 3)\nonumber
\eeqa
and the radial integrals give
\beqa
\cJ_{13}'&=&-\frac{\pi}{4}\left(\frac{}{}B(s,p)B(t,q)-B(1+s,p)B(1+t,q)\right)+(2\leftrightarrow 3)
\eeqa
The $\theta$-integral in $\cJ_{13}''$ gives
\beqa
\cJ_{13}''&=&2\pi\int_0^1dr_2\int_0^{r_2}dr_3\frac{[(r_2^2-r_3^2)+2(r_3^2-1)][2r_2^2(r_3^2-1)+(r_2^2-r_3^2)]\tK'}{r_2r_3(1-r_2^2)(1-r_3^2)(r_2^2-r_3^2)}+(2\leftrightarrow 3)\nonumber\\
&=&2\pi\int_0^1dr_2\int_0^{r_2}dr_3\left(\frac{2r_2}{r_3(1-r_2^2)}-\frac{1}{r_2r_3(1-r_2^2)}-\frac{1}{r_2r_3(1-r_3^2)}\right.\nonumber\\
&&\left.+\frac{4r_2}{r_3(r_2^2-r_3^2)}\right)+(2\leftrightarrow 3)\labell{J13}
\eeqa
The terms in the second  line  have no $p_2\inn p_3$ pole. Their radial integrals  can be evaluated using the formula \reef{II}, \ie
\beqa
\cJ_{13}''&\!\!\!\!\!=\!\!\!\!\!&\left(\frac{B(p,1+s+t)}{t}{}_3F_2\bigg[{1+s+t,\ t, \ -q \atop 1+s+t+p, \ 1+t}\ ;\ 1\bigg]-\frac{B(p,s+t)}{2t}{}_3F_2\bigg[{s+t,\ t, \ -q \atop s+t+p, \ 1+t}\ ;\ 1\bigg]\right.\nonumber\\
&\!\!\!\!\!\!\!\!\!\!&-\left.\frac{B(1+p,s+t)}{2t}{}_3F_2\bigg[{s+t,\ t, \ 1-q \atop 1+s+t+p, \ 1+t}\ ;\ 1\bigg]\right)\pi+(2\leftrightarrow 3)+J
\eeqa
The  term in the last line of \reef{J13} which we have called it $J$, however, is infinite when $r_2=r_3$ which means it has massless pole at $p_2\inn p_3$. So this part of $\cJ_{13}''$ must be calculated for $p_2\inn p_3\neq 0$.  

The $\theta$-integral in this part  has to be evaluated for $p_2\inn p_3\neq 0$. So we have to calculate the following integral:
\beqa
J&=&\int_0^1dx\int_0^1dy\,x^s y^{t-1}(1-x)^{p}(1-y)^{q}\int_0^{2\pi}d\theta (x+y-2\sqrt{xy}\cos(\theta))^{-1+p_2\cdot p_3}\nonumber
\eeqa
where we have chosen $x=r_2^2,\, y=r_3^2$. To take $\theta$-integral we use the following formula \cite{ISG}:
\beqa
\int_0^{2\pi}d\theta\frac{\cos(n\theta)}{(1+a^2-2a\cos(\theta))^b}&=&2\pi a^n\frac{\Ga(b+n)}{n!\Ga(b)}{}_2F_1\bigg[{b, \ n+b \atop n+1}\ ;\ a^2\bigg]
\eeqa
where $|a|<1$. One finds
\beqa
J&=&2\pi\int_0^1dx\int_0^xdy\,x^{-1+s+p_2\cdot p_3} y^{t-1}(1-x)^{p}(1-y)^{q}{}_2F_1\bigg[{1-p_2\inn p_3, \ 1-p_2\inn p_3 \atop 1}\ ;\ y/x\bigg]\nonumber\\
&&+2\pi\int_0^1dy\int_0^ydx\,x^{s} y^{-2+t+p_2\cdot p_3}(1-x)^{p}(1-y)^{q}{}_2F_1\bigg[{1-p_2\inn p_3, \ 1-p_2\inn p_3 \atop 1}\ ;\ x/y\bigg]\nonumber
\eeqa
 changing the variable  in the first line as $y=xu$ and in the second line as $x=yu$, one finds
\beqa
J&=&2\pi\int_0^1dx\int_0^1du\,x^{-1+p_2\cdot p_3+s+t} u^{t-1}(1-x)^{p}(1-xu)^{q}{}_2F_1\bigg[{1-p_2\inn p_3, \ 1-p_2\inn p_3 \atop 1}\ ;\ u \bigg]\nonumber\\
&&+2\pi\int_0^1dy\int_0^1du\,u^{s} y^{-1+s+t+p_2\cdot p_3}(1-yu)^{p}(1-y)^{q}{}_2F_1\bigg[{1-p_2\inn p_3, \ 1-p_2\inn p_3 \atop 1}\ ;\ u \bigg]\nonumber
\eeqa
These integrals have no massless pole in the open string $p$- or $q$-channel. So for ease of calculation we set $p=q=0$.  Then using the integral representation of the generalized hypergeometric function \reef{Fpq}
 one finds 
\beqa
J&=&2\pi\left(\frac{{}_3F_2\bigg[{t,\ 1-p_2\inn p_3, \ 1-p_2\inn p_3\atop 1+t, \ 1}\ ;\ 1\bigg]}{t(p_2\inn p_3+s+t)}+\frac{{}_3F_2\bigg[{1+s,\ 1-p_2\inn p_3, \ 1-p_2\inn p_3\atop 2+s, \ 1}\ ;\ 1\bigg]}{(1+s)(p_2\inn p_3+s+t)}\right)
\eeqa

Having found the explicit form of $\cJ_{13}$ and $\cJ_{14}$, one can use the package \cite{Huber:2005yg} to expand them. The result is
The low energy expansion of $\cJ_{13}$ and $\cJ_{14}$ are
\beqa
\cJ_{14}&=&\frac{\pi}{2}\left(\frac{1}{st}-\frac{\pi^2}{6}\bigg[\frac{q}{s}+\frac{p}{t}\bigg]+\cdots\right)\nonumber\\
\cJ_{13}&=&\frac{\pi}{2}\left(-\frac{3}{st}+\frac{\pi^2}{6}\bigg[-4+\frac{3p}{t}+\frac{3q}{s}\bigg]+\cdots\right)+J\nonumber
\eeqa
Note that these expansion for $\cJ_{14}$ and $\cJ_{13}-J$ are valid for the constraint \reef{kin}. The expansion for   $J$ is:
\beqa
J&=&2\pi\left(\frac{1}{p_2\inn p_3(s+t+p_2\inn p_3)}+\frac{1}{t(s+t+p_2\inn p_3)}-\frac{\pi^2}{6}+\cdots\right)\labell{88}
\eeqa
which is valid for $p=q=0$ but $p_2\inn p_3\ne 0$. 

We now check the relation \reef{iden33}. Using \reef{expand}, one finds the following expansion:
\beqa
2\cI_3p_1\inn N\inn p_2-2\cJ_1 p_2\inn V\inn p_2&=&-2\pi \left(\frac{1}{(s+t)}+\cdots\right)\nonumber
\eeqa
  for the case that $p_2\inn p_3=p=q=0$. Then the third relation in \reef{iden33} gives 
\beqa
\cJ_{13}p_2\inn p_3&=&2\pi \left(\frac{1}{(s+t)}+\cdots\right)\nonumber
\eeqa
which is consistent with the expansion \reef{88}.

\vspace*{1cm}

{\bf\Large {B. S-matrix element for $n=p+5$ case}}

In this appendix we consider the scattering amplitude \reef{A} for the case $n=p+5$. Since the  RR potential is totally antisymmetric, it must have at least four transverse indices. We consider the case that the RR potential carries four transverse indices and $p+1$ world volume indices. This is similar to the case $n=p-3$ and the RR potential with only world volumes indices which we studied in section 2.1. The cases that the RR potential carries more transverse indices is similar to the case $n=p-3$ and the RR potential with  transverse and world volumes indices which we studied in \cite{Garousi:2010bm}. 

Using the same steps as in section 2.1, one finds the scattering amplitude is zero for two graviton vertex operators. We find this result by explicit calculation in $(-3/2,-1/2)$-picture. One can find this result  in $(-1/2,-1/2)$-picture without explicit calculation, and by taking into account the fact that the RR field strength in the vertex operators is totally antisymmetric. 

The explicit calculation in $(-3/2,-1/2)$-picture, as we have done in section 2.1, gives the following result for two B-fields: 
\beqa
\cA&\!\!\!\!\!\sim\!\!\!\!\!&\frac{1}{4(p+1)!}\eps^{a_0\cdots a_p}{\veps _1}_{a_0\cdots a_p i j k l}{\veps_3}^{k l}\bigg( p_2^i p_3^j p_3\inn V\inn\veps_2\inn N\inn p_1  \cI_2-p_2^i p_3^j {p_3\inn N\inn\veps_2\inn N\inn p_1}  \cI_3 \nonumber\\
&&\left.
- p_2^i p_2\inn V\inn p_3 {p_3\inn V\inn\veps_2}^j \cJ_6- p_2^i {p_3\inn V\inn p_3} {p_3\inn V\inn\veps_2}^j \cJ_4
+ p_3^i{p_2\inn V\inn p_2}  {p_3\inn V\inn\veps_2}^j \cJ_2\right. \nonumber\\
&&\left.
 -{2}p_3^i{p_2\inn V\inn p_3} {p_3\inn V\inn\veps_2}^j \cJ_8-{2}p_2^i {p_2\inn N\inn p_3} {p_3\inn V\inn\veps_2}^j \cJ_9
-p_3^i{p_2\inn N\inn p_3} {p_3\inn V\inn\veps_2}^j \cJ_6\right. \nonumber\\
&&\left.
  -{2}p_2^i {p_3\inn V\inn p_3} {p_1\inn N\inn\veps_2}^j \cI_4-p_3^i {p_2\inn V\inn p_3}{p_1\inn N\inn\veps_2}^j \cI_2
-{2} p_2^i{p_2\inn V\inn p_3}{p_3\inn N\inn\veps_2}^j \cJ_7\right. \nonumber\\
&&\left.
  + p_2^i {p_3\inn V\inn p_3}{p_3\inn N\inn\veps_2}^j \cJ_{12}-  p_3^i{p_2\inn V\inn p_2} {p_3\inn N\inn\veps_2}^j \cJ_1
- p_3^i{p_2\inn V\inn p_3} {p_3\inn N\inn\veps_2}^j \cJ_6\right. \nonumber\\
&&\left.
 +p_3^i{p_2\inn N\inn p_3} {p_1\inn N\inn\veps_2}^j \cI_3 -p_2^i{p_2\inn N\inn p_3} {p_3\inn N\inn\veps_2}^j \cJ_6
-{2} p_3^i {p_2\inn N\inn p_3} {p_3\inn N\inn\veps_2}^j J_{10}\right. \nonumber\\
&&\left.
 +\frac{1}{2} p_2\inn V\inn p_2 p_3\inn V\inn p_3  {\veps_2}^{i j} \cJ_3+ \frac{1}{4} (p_2\inn V\inn p_3)^2{\veps_2}^{i j}\cJ_6+ \frac{1}{2} p_2\inn V\inn p_3 p_2\inn N\inn p_3 {\veps_2}^{i j} \cJ_{11}\right. \nonumber\\
&&\left.
 +\frac{1}{4} (p_2\inn N\inn p_3)^2 {\veps_2}^{i j}\cJ_6\bigg)\right. \nonumber\\
&&\left.+\frac{1}{2(p+1)!}\eps^{a_0\cdots a_p}{\veps _1}_{a_0\cdots a_p i j k l}p_2^i p_3^j\bigg(-{p_1\inn N\inn\veps_2}^k {p_1\inn N\inn\veps_3}^l \cI_1+ {p_1\inn N\inn\veps_2}^k {p_2\inn N\inn\veps_3}^l \cI_2\right. \nonumber\\
&&
- {p_1\inn N\inn\veps_2}^k {p_2\inn V\inn\veps_3}^l \cI_3\bigg)+(2\leftrightarrow 3)
\eeqa
The integrals are those  appear in \reef{total}.

We have argued before that the sum of  amplitudes \reef{gg} and  \reef{total} is invariant under linear T-duality when the world volume Killing coordinate is an index of the RR potential. However, it is  not invariant under linear T-duality if one of the indices of B-field/graviton polarization tensor which is contracted with the volume form, is the Killing coordinate. In that case one should add new amplitude  involving higher RR potential to make a complete T-dual amplitude \cite{Garousi:2009dj}. In this way one would find new amplitude for RR potentials $C^{(p-1)}$, $C^{(p+1)}$, $C^{(p+3)}$ and  $C^{p+5}$. The  $C^{p+5}$ part is exactly the terms in the seventh and eighth line above. All other terms involving $p_2^i$ or $p_3^i$ are not remnant of the amplitude \reef{total} under T-duality. The string theory produce them for other consistencies. 

The amplitude in terms of $H$ is
\beqa
\cA&\!\!\!\!\!\sim\!\!\!\!\!&\frac{1}{8(p+1)!}\eps^{a_0\cdots a_p}{\veps _1}_{a_0\cdots a_p i j k l}\bigg(\frac{2}{3}H_3^{klj}H_2^{ami}(p_3)_a(p_1)_m\cI_2-\frac{2}{3}H_3^{klj}H_2^{nmi}(p_3)_n(p_1)_m\cI_3 \nonumber\\
&\!\!\!\!\!\!\!\!\!\!&-H_3^{nlj}H_2^{mki}(p_1)_n(p_1)_m \cI_1+ H_3^{nlj}H_2^{mki}(p_2)_n(p_1)_m \cI_2-H_3^{alj}H_2^{mki}(p_2)_a(p_1)_m \cI_3\bigg)+(2\leftrightarrow 3)\nonumber
\eeqa
Since the integrals $\cI_1,\, \cI_2$ and $\cI_3$ have no constant and no massless open string poles \reef{expand}, the above amplitude does not produce any coupling for $C^{(p+5)}$. 

\end{document}